%% file: Main.tex
\documentclass[journal,comsoc]{IEEEtran}
\usepackage[table]{xcolor}

\usepackage[T1]{fontenc}

\ifCLASSINFOpdf
\else
\fi

\usepackage{amsmath}
\usepackage{bm}
\usepackage[cmintegrals]{newtxmath}

\usepackage{balance}

\usepackage[caption=false]{subfig}
\usepackage{graphicx}
\usepackage{subfig}
\usepackage{float}
\usepackage{url}
\usepackage{tabularx}
\usepackage{tikz}
\usepackage[hidelinks]{hyperref}
\usepackage{adjustbox}
\usepackage{booktabs}
\usepackage{rotating}
\usepackage{slashbox}
\usepackage{xcolor}
\usepackage{color}
\usepackage{colortbl}
\usepackage{multirow}
\usepackage[misc,geometry]{ifsym}
\usepackage{dirtytalk}
\usepackage[normalem]{ulem}

\usepackage{svg}
\usepackage{amsmath} 
\usepackage{physics} 
\usepackage{blindtext}
\usepackage[ruled,vlined]{algorithm2e}
\newcommand\Tstrut{\rule{0pt}{2.6ex}}       
\newcommand\Bstrut{\rule[-1.5ex]{0pt}{0pt}} 
\newcommand{\TBstrut}{\Tstrut\Bstrut} 
\definecolor{grey}{RGB}{197,197,197}
\newcommand{\orcid}[1]{\href{https://orcid.org/#1}{\includesvg[width=10pt]{orcid}}}
\usepackage{color, colortbl}
\definecolor{grey}{gray}{0.9}
\usepackage{cite}
\usepackage{lineno}
\usepackage[bottom]{footmisc}
\colorlet{mygreen}{green!60!gray}

\usepackage{tablefootnote}
\usepackage{threeparttable}

\begin{document}

\title{Demystifying the Transferability of Adversarial Attacks in Computer Networks}

\author{
Ehsan~Nowroozi ,~\IEEEmembership{Member,~IEEE,}~Yassine~Mekdad ,~\IEEEmembership{Member,~IEEE,} Mohammad Hajian Berenjestanaki, \\Mauro~Conti, ~\IEEEmembership{Fellow~Member,~IEEE,} and~Abdeslam~EL~Fergougui
\IEEEcompsocitemizethanks{
\IEEEcompsocthanksitem E. Nowroozi is with the Faculty of Engineering and Natural Sciences (FENS), Sabanci University, Istanbul, Turkey; and affiliated with the Center of Excellence in Data Analytics (VERİM), Sabanci University, Istanbul Turkey 34956; and affiliated with the Security \& Privacy Research Group (SPRITZ), University of Padua, 35121, Padua, Italy (e-mail: ehsan.nowroozi@sabanciuniv.edu, nowroozi@math.unipd.it).
\IEEEcompsocthanksitem Y. Mekdad is with the Laboratory of Computer Networks and Systems, Moulay Ismail University of Meknes, 11201, Zitoune, Meknes, Morocco, and with the Cyber-Physical Systems Security Lab, Department of Electrical and Computer Engineering, Florida International University, Miami, FL 33174 USA
 (e-mail: y.mekdad@edu.umi.ac.ma, ymekdad@fiu.edu).
\IEEEcompsocthanksitem M. Conti and M. Hajian Berenjestanaki are with the Department of Mathematics, University of Padua, 35121, Padua, Italy (e-mail: conti@math.unipd.it; mhajianb.1985@gmail.com).
\IEEEcompsocthanksitem A. El Fergougui is with the Laboratory of Computer Networks and Systems, Moulay Ismail University of Meknes, 11201, Zitoune, Meknes, Morocco (e-mail: a.elfergougui@umi.ac.ma). \\ 
*The corresponding authors of the paper are: Ehsan Nowroozi and Yassine Mekdad.
}
}
\maketitle
\input{Abstract}

\input{Introduction}

\input{Background}
\input{Related_Work}

\input{Methodology}

\input{Results_And_Discussion}

\input{Conclusion}
\input{Acknowledgment}

\bibliographystyle{IEEEtran}
\bibliography{References}
\begin{IEEEbiography}
[{\includegraphics[width=1in,height=1.25in]{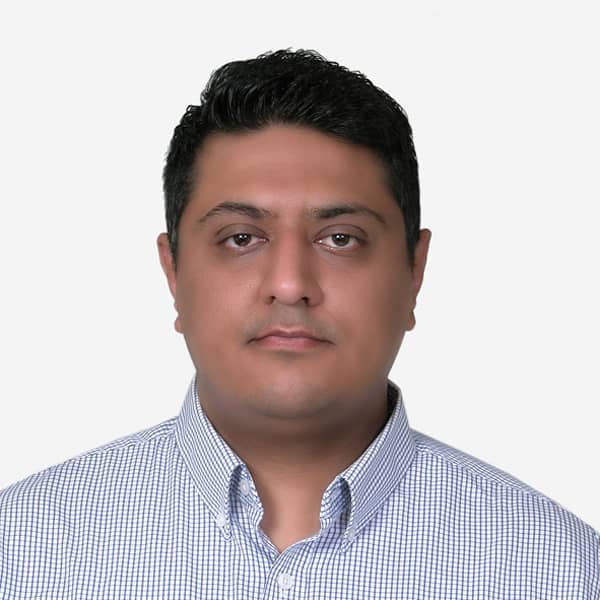}}]
{Ehsan Nowroozi} received his doctorate at the University of Siena in Italy. He is a postdoctoral researcher at Sabanci University's Faculty of Engineering and Natural Sciences (FENS), Istanbul, Turkey's Center of Excellence in Data Analytics (VERIM). In the years 2020 and 2021, he was a Postdoctoral Researcher at Siena University and Padova University in Italy. His principal research interests are in the fields of security and privacy, with a focus on the use of image processing algorithms in multimedia authentication (multimedia forensics), network and internet security, and other areas. His professional service and activity as a reviewer for IEEE TNSM, IEEE TIFS, IEEE TNNLS, Elsevier journal Digital, EURASIP Journal on Information Security.

\end{IEEEbiography}

\vskip -3\baselineskip plus -2fil
\begin{IEEEbiography}[{\includegraphics[width=1in,height=1.25in]{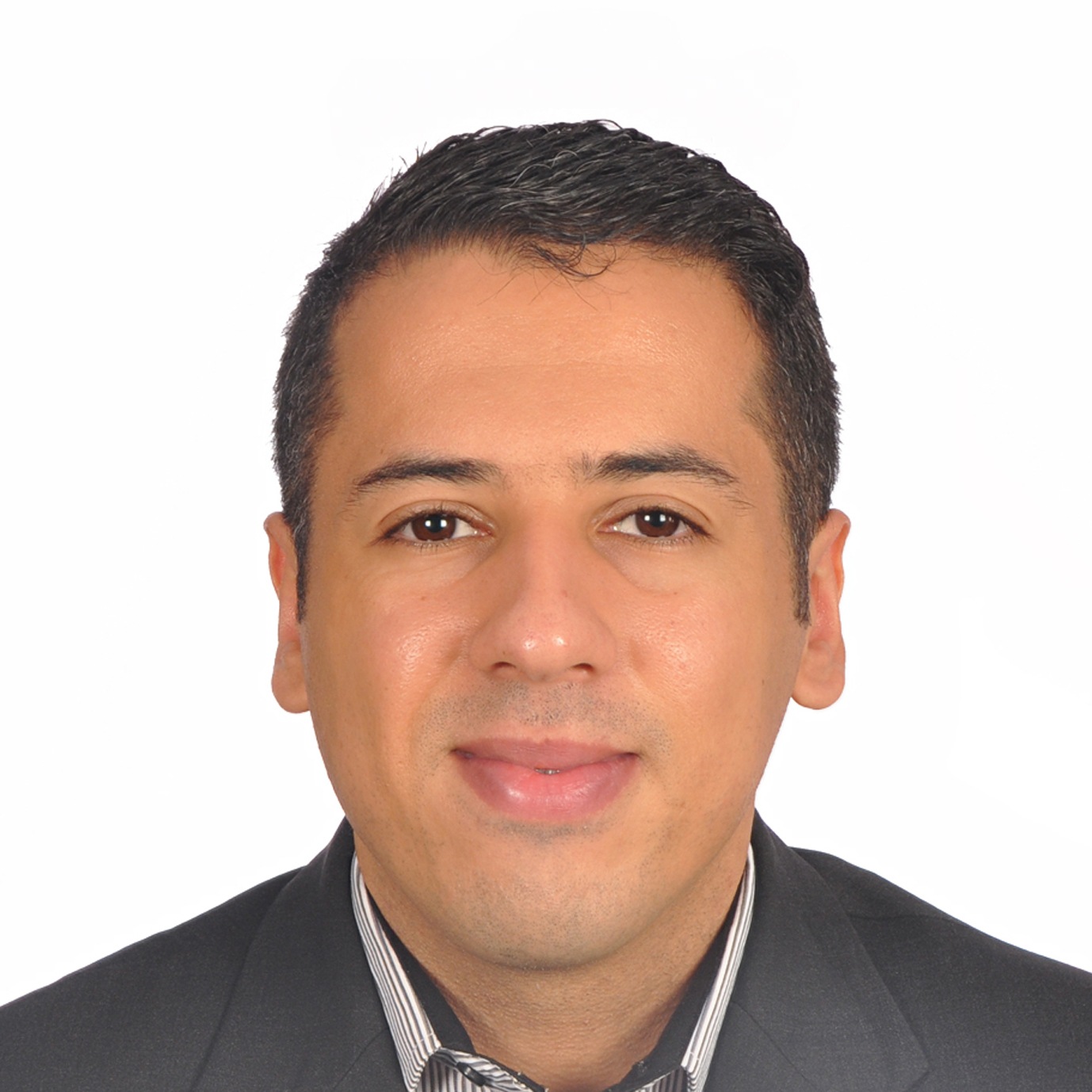}}]{Yassine Mekdad}
received a Msc Degree in Cryptography and Information Security from Mohammed V University of Rabat, Morocco. He works with the Computer Networks and Systems Laboratory at Moulay Ismail University of Meknes, Morocco. He holds a guest researcher position with the SPRITZ research group at University of Padua, Italy. He is currently working as a Research Scholar at the Cyber-Physical Systems Security Lab (CSL) at Florida International University, Miami, FL, USA. His research interest principally cover security and privacy problems in the Internet of Things (IoT), Industrial Internet-of-Things (IIoT), and Cyber-physical systems (CPS).
\end{IEEEbiography}

\vskip -2\baselineskip plus -1fil

\begin{IEEEbiography}[{\includegraphics[width=1in,height=1.25in]{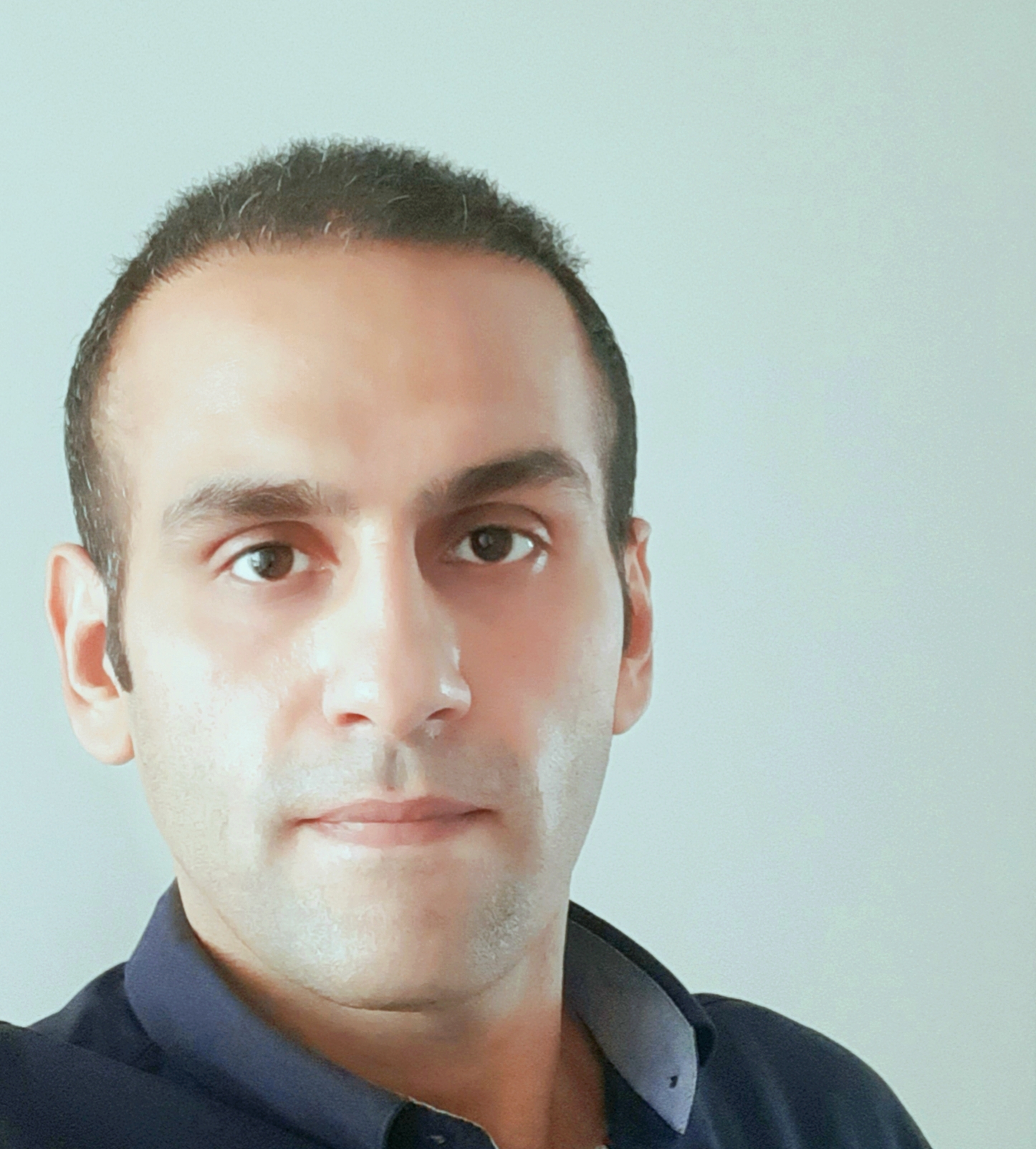}}]{Mohammad Hajian Berenjestanaki} is currently a research assistant at the SPRITZ security and privacy research group, University of Padua, Italy. He completed his bachelor’s degree in Electrical Engineering, Electronics at University of Mazandaran, Iran. He did his Master’s degree in Electrical Engineering, Communications, Cryptography in 2016 at the University of Tehran. His areas of interests include Networks Security and Cryptography, Deep Learning and Machine learning, Block-chain and Distributed Ledger Technology. He conducts his new researches in applied machine learning in Cybersecurity.
\end{IEEEbiography}

\vskip -2\baselineskip plus -1fil

\begin{IEEEbiography}[{\includegraphics[width=1in,height=1.25in]{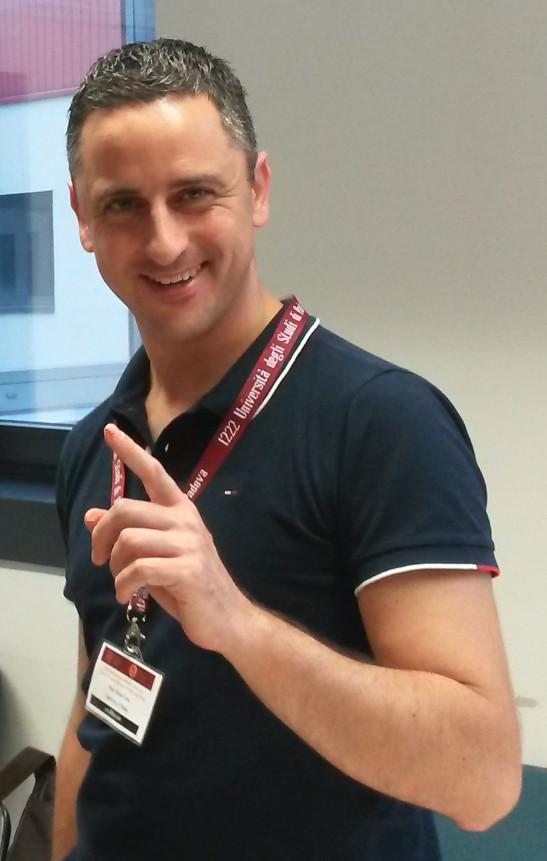}}]{Mauro Conti} received the
Ph.D. degree from the Sapienza University of Rome,
Italy. He is a Full Professor with the University
of Padua, Italy, and an Affiliate Professor with TU
Delft and University of Washington, Seattle. He
was a Postdoctoral Researcher with Vrije niversiteit
Amsterdam, The Netherlands. In 2011, he joined as
an Assistant Professor with the University of Padua,
where he became an Associate Professor in 2015,
and a Full Professor in 2018. He has been a Visiting
Researcher with GMU, UCLA, UCI, TU Darmstadt,
UF, and FIU. His research is also funded by companies, including Cisco,
Intel, and Huawei. He has been awarded with a Marie Curie Fellowship by
the European Commission, and with a Fellowship by the German DAAD.
\end{IEEEbiography}

\vskip -2\baselineskip plus -1fil
\begin{IEEEbiography}[{\includegraphics[width=1in,height=1.25in]{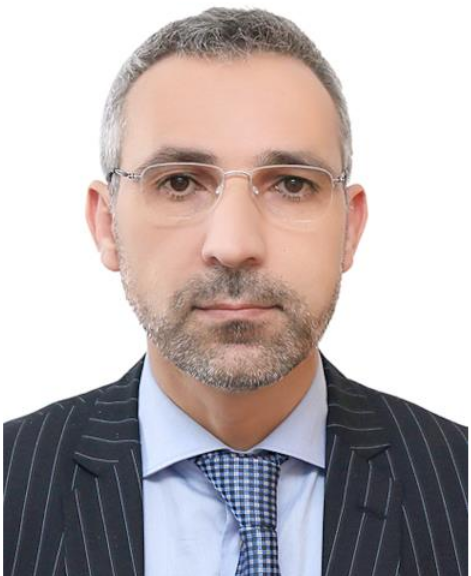}}]{Abdeslam El Fergougui}
received the PhD degree in Computer Science from Mohammed V University in Rabat - Morocco. He is currently a professor in the department of Computer science, Faculty of sciences Meknes, Moulay Ismail University. His current research interests include Network, Sensor Network, cryptography, security. Furthermore, he assured several advanced training workshops in its field of action. He also assured several training courses in several Moroccan and international universities. He has managed several national and international projects. He is an active member of AUF (University Agency of Francophonie) where he provides training at the international level. He is a member of the reading committee of several journals and conferences in his field.
\end{IEEEbiography}

\end{document}

%% file: Abstract.tex
\begin{abstract}

Convolutional Neural Networks (CNNs) models are one of the most frequently used deep learning networks, and extensively used in both academia and industry. Recent studies demonstrated that adversarial attacks against such models can maintain their effectiveness even when used on models other than the one targeted by the attacker. This major property is known as transferability, and makes CNNs ill-suited for security applications.

In this paper, we provide the first comprehensive study which assesses the robustness of CNN-based models for computer networks against adversarial transferability. Furthermore, we investigate whether the transferability property issue holds in computer networks applications. In our experiments, we first consider five different attacks: the Iterative Fast Gradient Method (I-FGSM), the Jacobian-based Saliency Map (JSMA), the Limited-memory Broyden Fletcher Goldfarb Shanno BFGS (L-BFGS), the Projected Gradient Descent (PGD), and the DeepFool attack. Then, we perform these attacks against three well-known datasets: the Network-based Detection of IoT (N-BaIoT) dataset, the Domain Generating Algorithms (DGA) dataset, and the RIPE Atlas dataset. Our experimental results show clearly that the transferability happens in specific use cases for the I-FGSM, the JSMA, and the LBFGS attack. In such scenarios, the attack success rate on the target network range from 63.00\% to 100\%. Finally, we suggest two shielding strategies to hinder the attack transferability, by considering the Most Powerful Attacks (MPAs), and the mismatch LSTM architecture.  
\end{abstract}


%% file: Introduction.tex
\section{Introduction}

\IEEEPARstart{I}{n} the 21st century, the evolution of artificial intelligence has demonstrated the tremendous benefit of deep learning algorithms to solve the most challenging problems of our society~\cite{Dong2021AApplications}. For example, applying deep learning techniques in the medical field shows promising results to perfectly detect and classify cancerous tissues through histopathology images~\cite{Sirinukunwattana2016LocalityImages}. Nowadays, existing Neural Networks (NNs) solutions have achieved human-level performance. In particular, Convolutional Neural Networks (CNNs) models which are among the most popular neural network architectures. Convolutional Neural Networks (CNNs) are increasingly employed in various computer network applications, such as network threat detection, network intrusion detection \cite{Vinayakumar2017}, malware detection in 5G-IoT Healthcare Applications \cite{Anand2021}, and multimedia forensics to detect several distorted pictures \cite{Barni2018Cnn-basedPost-processing}. 
Prior studies investigated the robustness of Graph Neural Networks against adversarial attacks\cite{zugner2020adversarial}, and demonstrated the possibility of successfully identifying important patterns of adversarial attacks against such networks. In this line of research, the fundamental issue is that even when only partial information of the graph is provided, or the attack is limited to a few perturbations, the classification performance constantly degrades. Furthermore, the attacks apply to a variety of node categorization schemes. As a result, the CNN model outperforms Graph Neural Networks regarding classification performance.

On the other hand, recent research studies have shown that some examples can be precisely computed to mislead machine-learning models \cite{Nowroozi2021AForensics}. To better illustrate these examples, Grosse et al.~\cite{Grosse2017AdversarialDetection} proved the possibility of crafting Android malware examples to mislead neural network malware detection models. The authors adapted the algorithm proposed by Papernot et al.~\cite{Papernot2016TheSettings} to craft the adversarial samples. More specifically, the proposed crafting algorithm utilizes the Jacobian matrix of the neural network.

To elaborate more, the generation of these examples are so-called adversarial examples and pose real-life security threats~\cite{Zhang2020AdversarialChallenges}. A real-world scenario consists of making a self-driving car misclassify the red light as green. This scenario could be a life-threatening example for autonomous vehicles. The popularity and diversity of the CNN applications make them a perfect target for adversarial examples. The concept of adversarial examples was introduced in 2014 by Szegedy et al. \cite{Szegedy2013IntriguingNetworks}. Later, many studies demonstrated the use of different adversarial attacks to compromise CNN-based models (e.g., offline handwritten signature verification, synthetic aperture radar image classification, object detection, automatic speech recognition). These attacks pose new security threats for systems built on machine learning algorithms, hindering the model's functionality in various applications.

It is worth mentioning that building adversarial examples can typically be applied to every machine learning algorithm due to the native characteristics of machine learning architectures. This vulnerability also affects modern deep neural networks. An interesting case in a black-box setting is that two models with different network architectures and trained with different training sets can misclassify the same adversarial example. In this scenario, the adversary trains the source network, and designs the adversarial example, then \textit{transfer} it to the victim's network even with a few knowledge about the victim's model \cite{PapernotTransferabilitySamples}. This substantial property is known as \textit{transferability} and might disrupt CNN's operation in various applications. To better visualize this concept, Wu et al.~\cite{WuUNDERSTANDINGEXAMPLES} evaluated the adversarial transferability in image classification against three datasets (MNIST, CIFAR-10, and ImageNet) and provided three key factors influencing it: the architecture, the test accuracy, and the model capacity. For the architecture, the similarity between the source and target model significantly enables the transferability property. Regarding the test accuracy, the models with high accuracy manifest high transferability. However, the models with deeper capacity (i.e., higher amount of model parameters) are less transferable when they act as source models.\\
\indent The purpose of our study is to see if and to what degrees the transferability of adversarial examples applies in computer networks, following some recent studies demonstrating that adversarial examples jeopardize CNN forensics techniques \cite{Marra2018OnIdentification}. 
From a research point of view, it is highly important to answer this question because the attack transferability would make developing defense strategies much more complex. Furthermore, giving the attacker a complete access to the source network will not prevent fooling the target network. Experimental evidence demonstrated by Tram\`er et al.~\cite{TramerTheExamples} has shown that the transferability of adversarial examples from a source network to a target network is more likely to occur in very large adversarial subspaces (i.e., the space of adversarial examples), where the two models will intersect. Consequently, the subspace dimensionality of two adversarial subspaces correlates with the transferred property. 
\\
\indent In this paper, we solely consider two instances of CNN's deep learning networks, which have lately been employed in computer vision, multimedia forensics, adversarial machine learning, and other fields. We mainly investigate if the attack transferability might occur in security-oriented applications for computer networks. In other words, we verify if an adversarial attack on a Source Network (SN) might be effective against a possibly different and unknown Target Network (TN). Although authors in \cite{Marra2018OnIdentification, PapernotTransferabilitySamples} recently demonstrated that the transferability of adversarial samples holds in computer vision applications and computer forensics. We launched five different attacks against three well-known datasets under black-box settings. Our experimental results demonstrate that the generated adversarial examples are not transferable in most cases. However, just a few cases of attacks are transferable from the SN to the TN. In particular, in the cross model transferability scenario where the same datasets such as N-BaIoT, DGA, and RIPE Atlas datasets are trained on two different network architectures. These adversarial examples achieve good transferability and can perfectly deceive the TN.
\subsection{Contributions}
Our contributions are summarized as follows:
\begin{itemize}

    \item We offer the FAIL models SPRITZ1 and SPRITZ2 that fails against different adversarial attacks, and which provides a basic framework for simulating and analyzing realistic adversaries. Across multiple degrees of adversarial control, we assess the robustness of the networks in terms of attack transferability against machine learning systems.
    
    \item We perform five different adversarial attacks with different parameters from low to high to demystify the transferability in computer networks, namely: the I-FGSM attack, the JSMA attack, the L-BFGS attack, the PGD attack, and the DeepFool attack.
    
    \item We independently examine the impact of training data and network architecture mismatch on the attacks' transferability. Furthermore, we systematically investigate the impact of partial adversarial knowledge and control on the robustness of machine learning models against the test-time attack (exploratory evasion-attack).
    
    \item We offer two shielding strategies that overcome prior adversarial attacks' limits. First, by using mismatch classifiers and considering the LSTM architecture as a target network to hinder the adversarial transferability. Second, by fine-tuning with Most Powerful Attacks (MPAs)~\cite{Barni2017MPA} to improve the security of the target network.

\end{itemize}

\subsection{Organization}
We outline the rest of our paper as follows. In Section~\ref{Background}, we provide the background and preliminaries on the datasets and adversarial attacks used in our experiments. Then, we overview the related works that discussed the transferability property in Section~\ref{Related}. Afterward, in Section~\ref{Methodology}, we provide our methodology and the experimental setup. Section~\ref{Results} presents the experimental results followed by a discussion evaluating the transferability in computer networks, and presents two recent approaches for hindering adversarial transferability. Finally, we conclude our paper and provide promising future research work in Section~\ref{Conclusions}.

%% file: Background.tex
\section{Background}
\label{Background}
In this section, we provide the background on the datasets and adversarial attacks used in our experiments. First, we present the N-BaIoT~\cite{nbaiot}, the DGA~\cite{Upadhyay2020}, and RIPE Atlas\cite{staff2015ripe} datasets. Afterward, we discuss the adversarial attacks performed to evaluate the transferability property in CNN-based models on computer networks. Table~\ref{Abbreviations} lists all the abbreviations and notations used in our paper.    

\input{Tables/Abbreviations}

\subsection{Datasets}
To evaluate the effectiveness of the transferability property in computer networks, we considered comprehensive real-world datasets that contain malicious and benign data in computer networks. It is worth mentioning the importance of relying on the high quality of datasets to achieve high training accuracy \cite{Tahaei2020TheSurvey}. In our experiments, we train the SN and the TN on three different datasets, namely: the N-BaIoT, the DGA, and the RIPE Atlas datasets. 
Additionally, these datasets are well-known and have lately been used for various cybersecurity purposes. The N-BaIoT dataset has been used to recognize botnet attacks in IoT devices employing deep auto-encoders \cite{Meidan2018N-BaIoT-Network-basedAutoencoders}, detection methods in the Internet of Things \cite{Abbasi2021}, and federated learning for detecting attacks in the Internet of things \cite{Rey2021FederatedDevices}. Besides, the DGA dataset is one of the most widely used datasets in detecting and classifying DGA botnets, and their families \cite{tuan2022detecting}. It presents a framework to identify Domain Generating Algorithms (DGAs) in real-world network traffic \cite{Upadhyay2020}. The RIPE Atlas dataset has been employed in JurisNN \cite{DECARVALHO2022108738}, network devices monitor and alerts \cite{sharma2022quic}, and so forth. In what follows, we provide a detailed description of each one of the considered datasets.

\subsubsection{\textbf{N-BaIoT datasets}}
The N-BaIoT datasets are built upon the collection of real benign and malicious traffic using port mirroring techniques and gathered from nine different commercial IoT devices. The N-BaIoT datasets contain 115 features and around seven million instances. With the increase of the IoT cyber attacks, which turns out the Internet of Things into the internet of vulnerabilities, the adversaries rely on using botnets to exploit such vulnerabilities \cite{Angrishi2017TurningBotnets}. The generation of malicious IoT traffic for the N-BaIoT datasets is performed by compromising nine commercial devices by two well-known IoT-based botnets: The \textit{BASHLITE} botnet and the \textit{Mirai} botnet. These botnets leverage the use of DDoS (Distributed Denial of Service) attack resulting in the unavailability of the IoT devices. The \textit{BASHLITE} botnet is a well-known IoT botnet that infects Linux-based IoT devices. With its capability of launching DDoS attacks, \textit{BASHLITE} infected up to 1 million IoT commercial devices, mostly DVRs and web-connected video cameras \cite{BASHLITEThreatpost}. The \textit{Mirai} botnet is one of the most popular IoT botnets that infected thousands of devices used in the past decades. The release of its source code shows that \textit{Mirai} botnet consists of (1) automatic scanning for vulnerable IoT devices, (2) uploading the malware into the vulnerable IoT device to perform the DDoS attack. The cyber attacks performed by the two authentic botnets in the IoT network cover flooding attacks (e.g., TCP-SYN flooding attack, UDP flooding attack), sending spam data, and scanning attacks.

The collection of malicious traffic generated by such attacks supports the adoption of N-BaIoT datasets in several applications. In \cite{Meidan2018N-BaIoT-Network-basedAutoencoders}, the authors provided a deep learning network-based anomaly detection method for IoT botnets. Their experimental results show that the use of deep autoencoders techniques can accurately detect IoT botnets. Another work proposed a privacy-preserving framework for malware detection in IoT devices \cite{Rey2021FederatedDevices}, the authors applied federated learning mechanisms for training and evaluating supervised and unsupervised learning models while protecting the privacy of sensitive shared data. We believe that we can also consider deep autoencoders approach utilized for IoT botnet detection to investigate the transferability property.
\subsubsection{\textbf{DGA datasets}}
The DGA datasets comprise legitimate and malicious DNs (Domain Names). The clean datasets are the Alexa (1M) records of top websites \cite{Upadhyay2020}, and the malicious DNs are gathered from more than 25 different families of malware (e.g.,  Kraken, nymaim, shiotob, and alureon) that exploit the DGA algorithm. These algorithms are embedded into the malware and principally designed to generate many malicious DNs automatically. In addition, DGA algorithms are resistant against blacklisting and sinkholing techniques, which detect and block malicious DNs. Integrating DGA algorithms into the malware enables C\&C (Command and Control) servers for botnets to perform potential DDoS attacks. 
Given the high quality and practical applications of DGA datasets, prior works in the literature have considered the DGA datasets. In~\cite{Luo2017DGASensor:Malwares}, the authors developed an approach to detect DGA-based malware. This approach enables the identification of a single DGA domain and relies on using a set of lexical evidence and RF (Random forest) algorithms to profile DGA-based malware. A similar work investigated the use of DL algorithms to detect DGAs \cite{Shahzad2021DGALearning}, the authors leveraged the RNN based architectures to detect potentially malicious DGAs, and their evaluation on real-world dataset demonstrated the high performance of RNN based DGA classifiers.

\subsubsection{\textbf{RIPE Atlas datasets}} The Internet of Things (IoT) concept has acquired much interest in past years. One approach to characterize IoT networks is through a large collection of Internet-aware detectors and controllers, including the electrical switches that turn the central air conditioning device on or off. The RIPE Atlas study assesses information about the Network through Internet-aware sensors~\cite{staff2015ripe}. RIPE Atlas could also be used to continually check the accessibility of networks or hosts from hundreds of locations across the world. Additionally, it can perform ad hoc connection tests to analyze and resolve identified network issues and test the availability of DNS servers. In these datasets, we considered feature-engineered RAW data by decoding the buffers returned in DNS queries, and returning the DNS measurement results~\cite{analysis}. Initially, the RIPE dataset aim to identify the application's type based on the flow data. These datasets are tested in two different network architectures: the Recurrent Neural Networks (RNN), and the Convolutional Neural Networks (CNN). We mention that several transformation steps have been performed to prepare the datasets~\cite{analysis2}, and the network is trained to classify the flow sequence for several applications. Although the accuracy of both training and testing data reached 80\%, the model can perfectly predict the majority of flow sequences, and the error rate is significantly reduced by feeding additional data into the system.
\subsection{Adversarial Attacks}
Much attention has been received recently concerning the possibility of adversarial examples to fool the CNN-based models. 
In the existing literature, the adversary can perform the adversarial attack under three different settings: (i) white-box, (ii) gray-box, and (iii) black-box attacks. (i) In the white-box setting, the adversary has a Perfect Knowledge (PK) on the TN by adopting a particular categorization for adversarial examples \cite{Zhang2020AdversarialChallenges}. (ii) In the gray-box setting, the adversary has a Limited Knowledge (LK) on the TN \cite{NEURIPS2018}. (iii) In the black-box setting, the attacker does not have access to the internal details of the TN, which is a more feasible and complicated scenario than the previous ones; as a result, the adversary uses multiple queries to obtain such internal details. Therefore, we are motivated to consider the black-box scenario given its feasibility in real-world applications. In our study, we propose and test a generalized technique for black-box threats on ML that takes advantages of adversarial transferability.

In our work, we performed five adversarial attacks under black-box settings, where we do not have access to the underlying deployed model: The I-FGSM attack \cite{Kurakin2017ADVERSARIALWORLD}, the JSMA attack~\cite{Papernot2016TheSettings}, the L-BFGS attack \cite{Szegedy2013IntriguingNetworks}, the PGD attack \cite{Madry2018}, and the DeepFool attack~\cite{Moosavi-DezfooliDeepFool:Networks}. 
In our experiments, we considered the Foolbox library \cite{FoolBox} to implement all adversarial attacks.
\hfill\\
\subsubsection{\textbf{The I-FGSM attack}}
The I-FGSM attack represents the iterative version of the FGSM (Fast Gradient Sign Method)~\cite{GoodfellowEXPLAININGEXAMPLES}. The FGSM attack is one of the most well-known adversarial examples. It consists of targeting neural networks, which fails the network and potential misclassification of an input. The Fast Gradient Sign Method is a gradient-based attack that adjusts the input data to maximize the loss. Given an input $X$ of a model with $\theta$ parameters and a ground truth label $Y$. We compute the adversarial example $Adv_{X}$ of the FGSM attack as follows:  
\begin{equation}
Adv_{X} = X + \epsilon \times sign(\triangledown_{X}J(\theta,X,Y)).
\end{equation}
Where $\epsilon$ is the normalized attack strength factor, and $J$ is the cross-entropy cost function used to train the model. We mention that $\epsilon$ is the parameter controlling the perturbations and should be considered minimal for the success of the adversarial example. Given that the FGSM attack often fails in some scenarios, the iterative version of the FGSM attack, namely the I-FGSM attack \cite{Kurakin2017ADVERSARIALWORLD}, guarantees finer perturbations and could be described as a straightforward extension of the FGSM. More specifically, the I-FGSM attack considers iterating the FGSM method with smaller steps on the way to the gradient sign. For the I-FGSM attack, we express the adversarial example for each $i+1$ iteration such that $Adv^{0}_{X} = X$ as follows: 
\begin{equation}
        Adv^{i+1}_{X} = Adv^{i}_{X} + \epsilon * sign(\triangledown_{X}J(\theta,Adv^{i}_{X},Y)).
\end{equation}
\subsubsection{\textbf{The JSMA attack}}
Papernot et al. \cite{Papernot2016TheSettings} introduced the JSMA attack to reduce the number of perturbations comparing to the I-FGSM attack. Therefore, the construction of JSMA-based adversarial examples is more suitable than I-FGSM for targeted misclassification and harder to detect. However, the JSMA algorithm requires a high computational cost. The JSMA attack is a gradient-based method that uses a forward derivative approach and adversarial saliency maps. Instead of a backward propagation like the I-FGSM attack, the forward derivative is an iterative procedure that relies on forward propagation. Given an $N$-dimensional input $X$ of a model, the adversary starts by computing a Jacobian matrix of the $M$-dimensional classifier $F$ learned by the model during the training phase. The Jacobian matrix is given by:
\begin{equation}
        J_{F}(X) = \pdv{F(X)}{X} = {\begin{bmatrix} 
        \pdv{F_{j}(X)}{x_i}.
        \end{bmatrix}}_{i \in 1..M, j \in 1..N}.
\end{equation}
Afterward, the adversary produces adversarial samples by constructing the adversarial saliency maps $S$ based on the forward derivative. These maps identify specific input features where the adversary enables the perturbations needed for the desired output. Thus, building the adversarial sample space. For a target class $t$, we define the adversarial saliency map by the following equation:
\begin{equation}
S(X, t)[i]=\left\{\begin{array}{ll} 0, if \frac{\partial F_{t}}{\partial x_{i}}(X)<0 \text { or } \sum_{j \neq t} \frac{\partial F_{j}}{\partial x_{i}}(X)>0. & \\ {\frac{\partial F_{t}}{\partial x_{i}}(X)\left|\sum_{j \neq t} \frac{\partial F_{j}}{\partial x_{i}}(X)\right|,}otherwise. & \end{array}\right.
\end{equation}
In summary, JSMA is a greedy repeated process that generates a saliency map at each iteration through forwarding propagation, indicating which pixels contributed the most to classification. A large value in this map, for instance, indicates that changing that pixel significantly increases the likelihood of selecting the incorrect class. The pixels are then changed individually, depending on this map, by a relevant number {$\theta$} ({$\theta$} refers to the range of possible pixel changes).
\hfill\\
\subsubsection{\textbf{The L-BFGS attack}}
The L-BFGS attack has been proposed by Szegedy et al. \cite{Szegedy2013IntriguingNetworks}. It is a non-linear gradient-based numerical optimization method and aims to reduce the perturbations for a given input. The L-BFGS attack generates the adversarial examples by optimizing the input $X$ and maximizing the prediction error. The L-BFGS attack is a practical box-constrained optimization method to generate adversarial samples. However, it is time-consuming and computationally demanding. For an input $X$, The formal expression of the optimization problem of the L-BFGS attack is given by:
\begin{equation}
\label{Eq}
\min_{Adv_{X}} || X - Adv_{X} ||^{2}_{2}  \mbox{  subject to: } F(X+r) = l.
\end{equation}
Such that $F$ is the classifier, $l$ is the target label, and $r$ is the minimizer. However, the optimization problem depicted in Eq.~\ref{Eq} is computationally expensive and considered as a hard problem. An approximate solution solves this problem by using a box-constrained L-BFGS \cite{Szegedy2013IntriguingNetworks}, and is expressed as follows:
\begin{equation}
\min_{Adv_{X}} c.|| X - Adv_{X} ||^{2}_{2} - J(\theta,X,Y). 
\end{equation}
In this approximation, the objective is to find the minimum scalar $c > 0$ and for which the minimizer $r$ of the Eq.~\ref{Eq} is satisfied.
\hfill\\
\subsubsection{\textbf{The PGD attack}}
The PGD attack \cite{Madry2018} is a \textit{first-order adversarial attack}, i.e., consists of using the local first-order information about the targeted network. It is an iterative version of the FGSM, such as the I-FGSM. Except that in the PGD attack, for each iteration $i+1$, the input $X$ is updated according to the following rule:
\begin{equation}
 Adv^{i+1}_{X} = \Pi_{X+S}( Adv^{i}_{X} + \epsilon * sign(\triangledown_{X}J(\theta,X,Y))).
\end{equation}
Where $\Pi$ is the projection operator that keeps $Adv^{i+1}_{X}$ within a set of perturbation range $S$. We note that the PGD attack looks for the perturbations that maximize $J(\theta,X,Y)$ while considering the $L^\infty$ distortion.
\hfill\\
\subsubsection{\textbf{The DeepFool attack}}
Regarding the DeepFool attack~\cite{Moosavi-DezfooliDeepFool:Networks}, it fools the binary and multiclass classifiers by finding the closest distance from the input $X$ to the other side of the classification boundary. For any given classifier, the minimal perturbation $r$ to generate an adversarial example by the DeepFool attack is formally expressed by:
\begin{equation}
 \delta(X,F) = \min_{r}||r||_{2} \mbox{ subject to: } F(X+r) \neq F(X). 
\end{equation}
Such that $\delta$ describes the robustness of $F$ at the input $X$. We mention that the state-of-the-art classifiers are weak against the DeepFool attack. Therefore, the DeepFool present a reference to test and evaluate the robustness and the performance of existing classifiers. 
The N-BaIoT datasets are built open the collection of real and benign malicious traffic in pcap using nine gathered 

%% file: Tables/Abbreviations.tex
\begin{table}[!h]
\centering
\caption{Abbreviation and Notation List. \label{Abbreviations}}
\begin{tabular}{@{}|c|l|}
\hline
\textbf{Acronym} & \textbf{Description} \\ \hline

CNN & Convolutional Neural Network \\ \hline

LSTM & Long Short-Term Memory Networks \\ \hline



LK & Limited Knowledge \\ \hline

PK & Perfect Knowledge \\ \hline

CF & Counter Forensics \\ \hline

DL, ML & Deep Learning, Machine Learning \\ \hline


SN, TN & Source Network, Target Network \\ \hline


CM  & Cross Model \\ \hline

CT  & Cross Training\\ \hline

CMT  & Cross Model and Training\\ \hline

ASR & Average Attack Success Rate \\ \hline

Max. dist & Average Maximum distortion\\ \hline

PSNR & Average Peak signal-to-noise ratio\\ \hline

$L_1$ dist & Average $L_1$ distance\\ \hline

FGSM & Fast Gradient Sign Method\\ \hline

I-FGSM & Iterative Fast Gradient Sign Method \\ \hline

JSMA  & Jacobian-based Saliency Map Attack\\ \hline

LBFGS  &  Limited-memory Broyden-Fletcher-Goldfarb-Shanno\\ \hline

PGD  & Projected Gradient Descent\\ \hline

N-BaIoT  & Network-based Detection of IoT\\ \hline

DGA  & Domain Generating algorithms\\ \hline

DN  & Domain Name\\ \hline

RNN & Recurrent Neural Network\\ \hline

DDoS  & Distributed Denial of Service\\ \hline



\end{tabular}
\end{table}

%% file: Related_Work.tex
\section{Related Work}
\label{Related}
The early studies of adversarial examples revealed the transferability of adversarial attacks \cite{GoodfellowEXPLAININGEXAMPLES,Szegedy2013IntriguingNetworks}. Recently, initial efforts have been devoted to explore this property in different areas (e.g., multimedia forensics, natural language processing).

In multimedia forensics applications, the use of machine learning techniques offers promising solutions for security analysts. However, adversarial image forensics significantly hinder such possibility \cite{Nowroozi2021AForensics}. Barni et al. \cite{Barni2019OnForensics} investigated the transferability of adversarial examples in CNN-based models. The authors launched the I-FGSM and the JSMA attacks against two different datasets and argued that adversarial examples are non-transferable under black-box settings. In the same context, another work explored the transferability and vulnerability of CNN-based methods for camera model investigations \cite{Marra2018OnIdentification}. Later, the authors in \cite{WangAdmix:Attacks} proposed an input transformation-based attack framework to enhance the transferability of adversarial attacks. 

In Natural Language Processing (NLP) applications, Wallace et al.~\cite{Wallace2020ImitationSystems} investigated the transferability in black-box Machine Translation (MT) systems through gradient-based attacks. In their proposed approach, the authors built an imitation model similar to the victim model. Next, they produced transferable adversarial examples to the production systems.

In Computer Vision (CV) applications, it is worth mentioning that deep learning-based adversarial attacks have become a significant concern for the artificial intelligence community \cite{Akhtar2018ThreatSurvey}. The transferability of adversarial examples has also been investigated in computer vision tasks \cite{Jia2019EnhancingReduction}. In particular, the authors in \cite{Jia2019EnhancingReduction} proposed adversarial examples to enhance the cross-task transferability in classification, object detection models, explicit content detection, and Optical Character Recognition (OCR).

In different application scenarios, Suciu et al. \cite{SuciuWhenAttacksb} implemented a framework to evaluate realistic adversarial examples on image classification, android malware detection, Twitter-based exploit prediction, and data breach prediction. Then, proposed evasion and poisoning attacks against machine learning systems. They prove that a recent evasion attack is less effective in the absence of broad transferability. The authors explored the transferability of such attacks and provided generalized factors that influence this property. In a similar work, Demontis et al. \cite{DemontisWhyAttacks} evaluated the transferability of poisoning and evasion attacks against three datasets that are associated with different applications, including android malware detection, face recognition, and handwritten digit recognition. The authors demonstrated that the transferability property strongly depends on three metrics: the complexity of the target model, the variance of the loss function, and the gradient alignment between the surrogate and target models. However, the considered metrics are not well proven in security-oriented applications and might add restrictions on the threat model. 

In Table~\ref{difference}, we present a comparison that summarizes our work and current works on the transferability in machine and deep learning applications. To the best of our knowledge, most covered research in this area did not consider the transferability property in computer networks. To that end, we aim to evaluate the robustness of transferable examples in CNN-based models for computer networks by performing a comprehensive set of experiments under black-box settings.
\input{Tables/Differences}

%% file: Tables/Differences.tex
\begin{table*}[!h]
\centering
\scriptsize
\caption{Difference between our work and existing works on the transferability in machine and deep learning applications \label{difference}}
\resizebox{2\columnwidth}{!}{
\begin{tabular}{|l | l | l | l | l | l |}
\hline
\textbf{Ref.} & \textbf{Application Domain} & \textbf{Adversarial Attacks} & \textbf{Considered Datasets} & \textbf{Model Settings} & \textbf{Transferability Results}   \\ \hline

    \cite{Barni2019OnForensics} & Multimedia forensics & -JSMA, and I-FGSM             &      -RAISE~\cite{dang2015raise}, and VISION~\cite{shullani2017vision}                  & -Black box &       -No transferability between\\ & & & & &   the SN and the TN                         \\ \hline
   
    \cite{Marra2018OnIdentification}    &    Multimedia forensics                &   -FGSM, and PGD                   &            -VISION~\cite{shullani2017vision}         &     -White box           &                   -No transferability between            \\ & & & & & the SN and the TN\\\hline
   
       \cite{WangAdmix:Attacks}      &      Image processing             &     -Input transformation         &     -ImageNet~\cite{russakovsky2015imagenet}                &    -White box    & -The transferability happens\\ & &(Admix) & & -Black box &     in white and black box settings                         \\

  \hline
  
   \cite{Wallace2020ImitationSystems}          &        Natural Language             &       -Gradient-based                &         -IWSLT~\cite{cettolo2014report}, and Europarl~\cite{koehn2005europarl}       &        -Black box                   & -The transferability happens, and   \\ 
   & Processing (NLP)       &    & & & a defense system is proposed\\
   
   \hline

       \cite{Jia2019EnhancingReduction}     &   Computer Vision (CV)                 &       -Dispersion reduction (DR)               &            -ImageNet~\cite{russakovsky2015imagenet}, NSFW Data         &    -Black box            &    -The adversarial transferability happens \\
       
        & & & Scraper~\cite{Scraper}, and COCO-Text~\cite{veit2016coco}  & &  on intermediate convolution layers                      \\
       
       \hline

    \cite{SuciuWhenAttacksb} &   Different applications                &       -The FAIL attack model              &   -Drebin Android malware detector~\cite{arp2014drebin},       &      -Black box          &         -Bypass the TN by considering                       \\ 
    & &        &Twitter-based exploit prediction~\cite{sabottke2015vulnerability},  & & the transferability property\\
    & & & and Data breach prediction~\cite{liu2015cloudy}& & \\
    
    \hline
    
\cite{DemontisWhyAttacks}          & Different applications            &        -Gradient-based Evasion            &    -MNIST89~\cite{deng2012mnist}, and Drebin~\cite{arp2014drebin}                 &    -Black box            &         -Different metrics that impact                         \\ 

& &  and Poisoning & & -White box &on attacks' transferability \\ 
\hline

    Our work      &          Computer Networks          &     -JSMA, I-FGSM, LBFGS,                &          -N-BaIoT~\cite{nbaiot}, DGA~\cite{Upadhyay2020}, and RIPE Atlas\cite{staff2015ripe}       &       -Black box         &           -The adversarial transferability                      \\ 

& &  PGD, and DeepFool & & & happens for the JSMA, I-FGSM, \\
 &  &  & & & and LBFGS attacks\\

\hline
\end{tabular}}
\end{table*}

%% file: Methodology.tex
\section{Methodology}
\label{Methodology}
To evaluate the transferability property of adversarial attacks against CNN-based models in IoT networks, which are trained with the N-BaIoT \cite{Meidan2018N-BaIoT-Network-basedAutoencoders}, DGA \cite{Upadhyay2020}, and RIPE Atlas \cite{staff2015ripe} datasets. We considered two network architectures namely \textit{SPRITZ1} and \textit{SPRITZ2}, and five different adversarial attacks (as described in Sect.~\ref{Background}). Given the datasets and the network architectures, we built six classes of networks as described in Table ~\ref{Networks}. In what follows, we present the architectural details of the network and the experiments carried out during our investigation.

\begin{figure}[!ht]
\begin{center}
\includegraphics[width=0.31\textwidth,height=4cm]{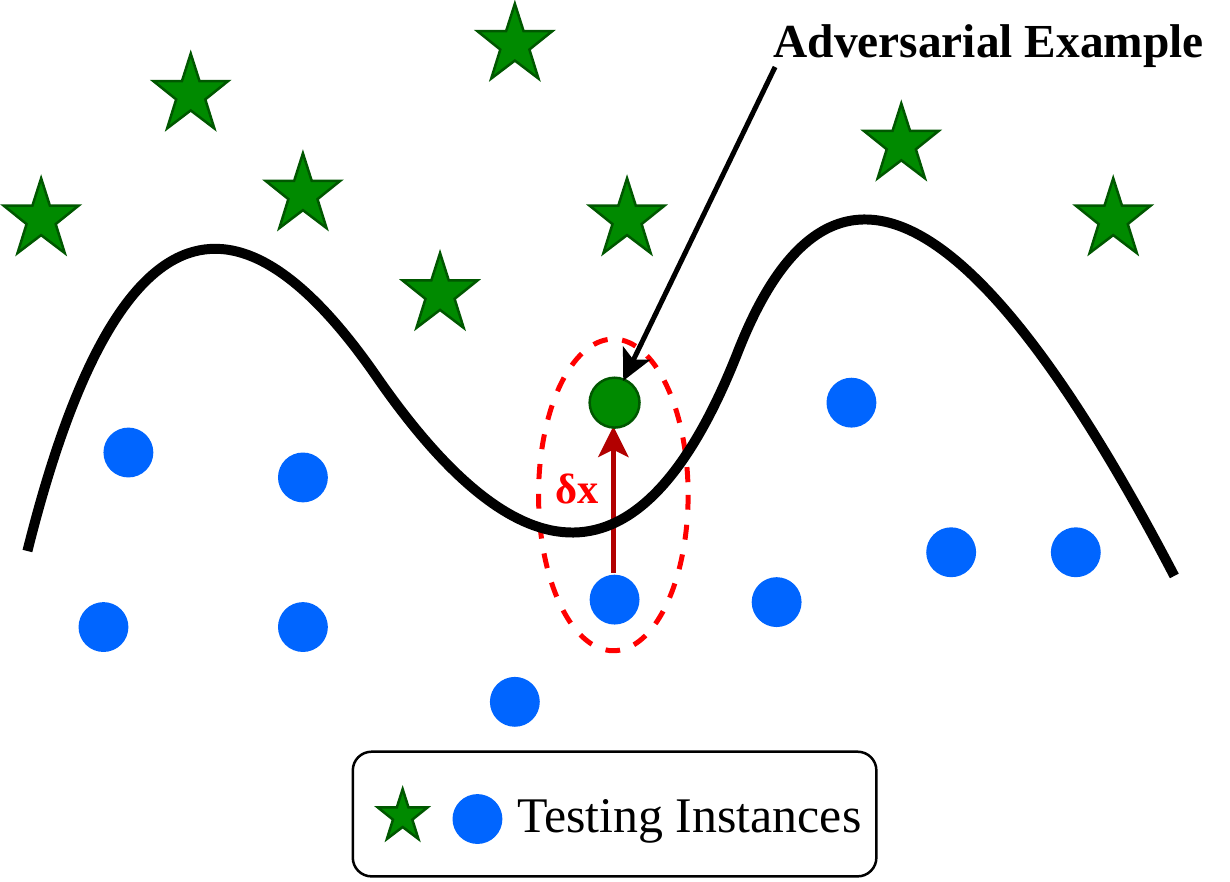}
\caption{Representation of adversarial attack against machine learning classifier. The objective is to generate a small perturbation $\delta x$ that would result the instance crossing the decision boundary~\cite{Evasion-Plot}.}
\label{Evasion}
\end{center}
\end{figure}

In ML and DL models, the adversary can influence the amount of training and testing data depending on his capability. As a result, the attacker's capability is relevant to the impact of the attack. In the \textit{exploratory-evasion attack} scenario, the adversary's capability is limited to alterations to the testing data, while changes to the training data are not permitted. In a \textit{causative or poisoning attack scenario}, the attack might disrupt the training phase; these attacks are sometimes referred to as poisoning attacks. The majority of the computer network attacks developed so far come within the group of exploratory attacks. In this study, we examined the exploratory evasion scenario \cite{nowroozi2020machine} in which the adversary's capabilities are limited to the alterations of the testing data, while the changes on the training data are not permitted. In the case of causative attacks, the attack might disrupt the training process, such as poisoning attacks that are commonly used to describe these types of attacks. The majority of the planned computer network attacks in ML and DL belong to the first scenario. These test time attacks operate by modifying the target sample and pushing it beyond the decision boundary of the model without changing the training process or the decision boundary itself, as shown in Figure \ref{Evasion}.
\subsection{Network Architecture}
In our experiments, we considered the network architecture provided by \cite{Barni2018Cnn-basedPost-processing}. It is a CNN-based network that includes an input layer, convolutional layers, max-pooling layers, and one fully connected layer. The network is characterized by several convolutional layers before the first max pooling layer to enable at a large scale a good reception for each neuron. Additionally, we maintain the maximum spatial information by setting the stride to 1. Accordingly, we deployed the \textit{SPRITZ1} and \textit{SPRITZ2} networks with different types of datasets (ref. Table ~\ref{Networks}). It is worth mentioning that we are considering the deep network given its difficulty to compromise compared to the shallow network. Moreover, we rely on such network architecture because it achieves a high accuracy during the training phase. Both networks are configured once as the SN and once as the TN. The architectural details of \textit{SPRITZ1} and \textit{SPRITZ2} networks are given below.
\input{Tables/Networks}
\hfill\\

\subsubsection{\textbf{SPRITZ1 Network}}
\begin{figure}[!ht]
\begin{center}
\includegraphics[width=0.43\textwidth,height=3.5cm]{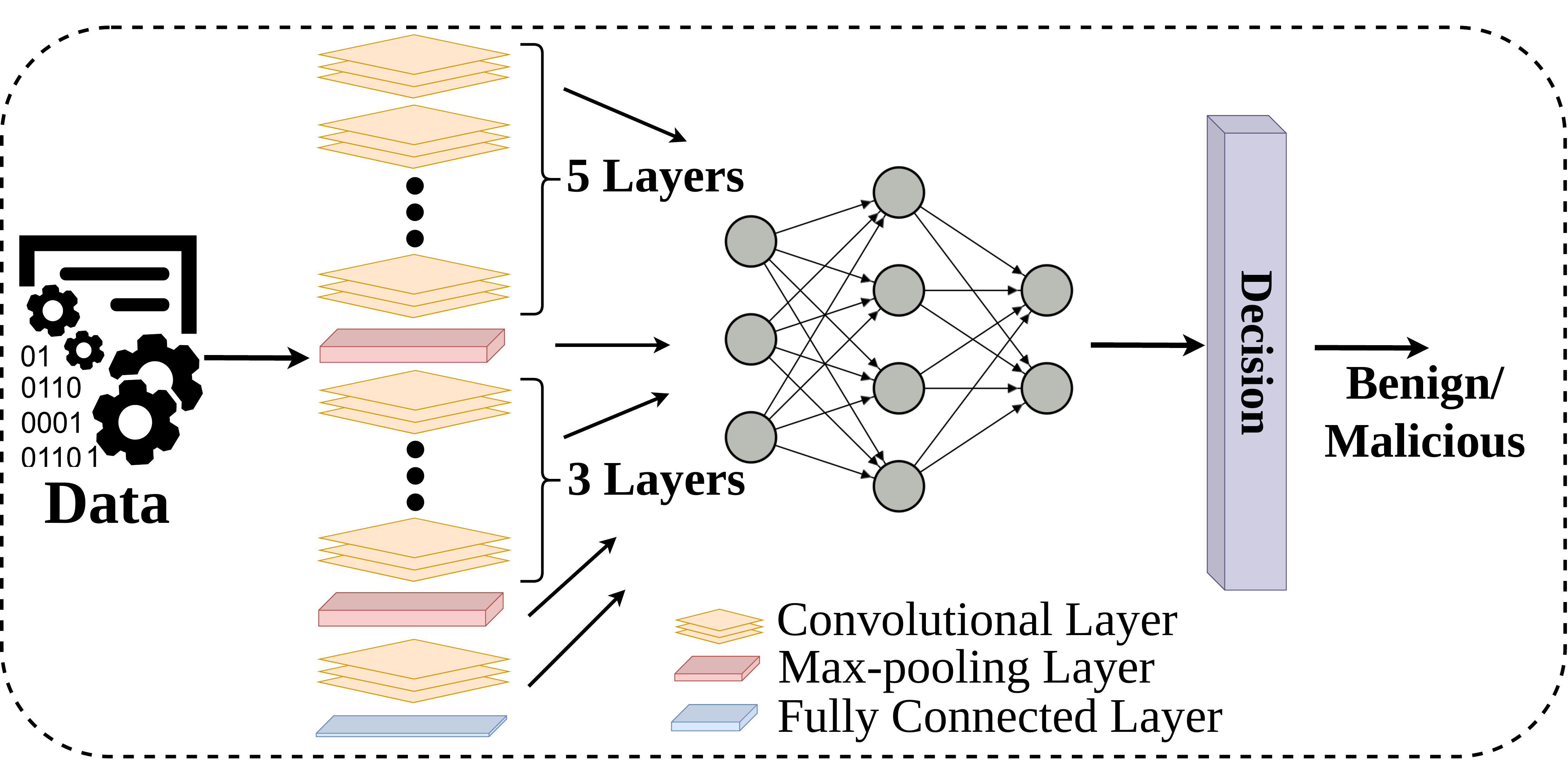}
\caption{Pipeline of the proposed \textit{SPRITZ1} architecture (shallow network).}
\label{Spritz1}
\end{center}
\end{figure}

The configuration of \textit{SPRITZ1} is a shallow network that consists of nine convolutional layers, two max-pooling layers, and one fully connected layer. In Figure ~\ref{Spritz1}, we illustrate the pipeline of the \textit{SPRITZ1} architecture. For all the convolutional layers, we use the kernel size of $3 \times 3$ and stride 1. In addition, we applied the max-pooling with the kernel size of $2 \times 2$ and stride 2. We note that by halving the number of features maps in the final convolutional layer, we reduce the number of parameters and consider only one fully-connected layer.

\subsubsection{\textbf{SPRITZ2 Network}}
\begin{figure}[!ht]
\begin{center}
\includegraphics[width=0.43\textwidth,height=3.5cm]{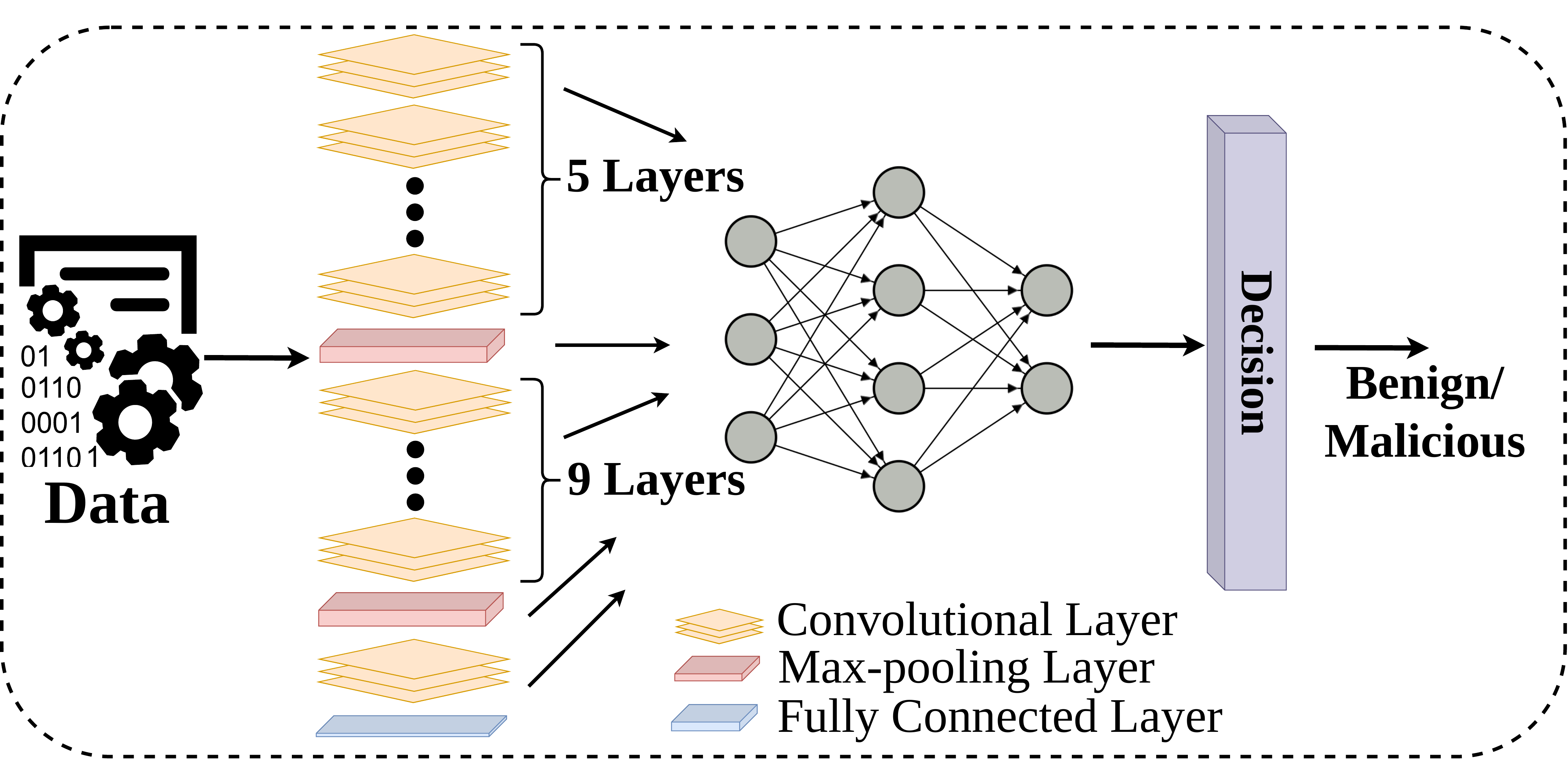}
\caption{Pipeline of the proposed \textit{SPRITZ2} architecture (deep network).}
\label{Spritz2}
\end{center}
\end{figure}

In the \textit{SPRITZ2} network, we increased the number of convolutional layers to fifteen layers comparing to \textit{SPRITZ1}. As depicted in Figure~\ref{Spritz2}, the \textit{SPRITZ2} network is deeper than \textit{SPRITZ1} network. However, we respected the same settings as the \textit{SPRITZ1} network. In other words, we used one fully connected layer and two max-pooling layers with the kernel size of $2 \times 2$ and stride 1. For all the convolutional layers, we considered the kernel size as $3 \times 3$ and stride 1.
\subsection{Experimental Setup} 

To build our models $N_{SPRITZ1}^{N-BaIoT}$ and $N_{SPRITZ2}^{N-BaIoT}$, for each class, we considered 58000 patches for training, 20000 for validation, and 20000 for testing, per class. The input patch size is set to $64 \times 64$ in all the cases. Regarding $N_{SPRITZ1}^{DGA}$ and $N_{SPRITZ2}^{DGA}$, we considered 2000 patches for training and 500 patches for validation and testing, per class. Concerning $N_{SPRITZ1}^{RIPE}$ and $N_{SPRITZ2}^{RIPE}$, we employed 60000 patches for training, 20000 for validation, and 20000 for testing, per class, that we think these number of features enough for a CNN for generalizing. By using small patches, we increase the depth of the network with the same parameters. Consequently, the detection accuracy is increased when aggregating the patch scores.
We implemented our CNN, and LSTM networks for training and testing in TensorFlow via the Keras API \cite{chollet2015} with Python. We ran our experiments using Intel\textregistered Core™ i7-10750H, NVIDIA\textregistered GeForce RTX™ 2060 with 6GB GDDR6 GPU, DDR4 16GB*2 (3200MHz), Linux Ubuntu 20.04. We made the simulation Python code publicly available at Github \cite{GitHub}. The number of training epochs is set to 20. We used Adam solver with a learning rate of $10^{-4}$ and a momentum of 0.99. The batch size for training and validation on N-BaIoT, DGA, and RIPE is set to 64, and the test batch size is set to 100. 

\subsection{Empirical Study}

\begin{figure}[!ht]
\begin{center}
\includegraphics[width=0.43\textwidth,height=3.5cm]{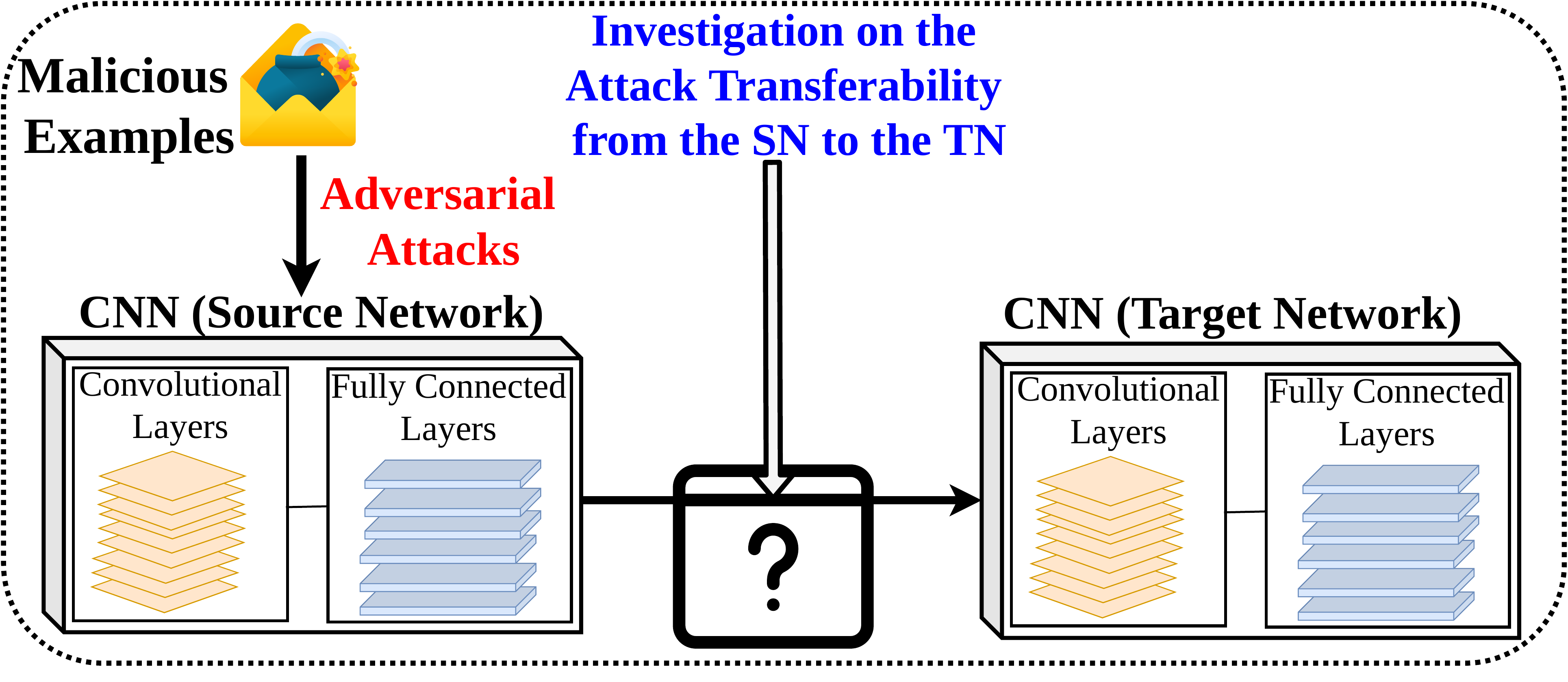}
\caption{General scheme of the transferability property in computer networks.}
\label{transferability}
\end{center}
\end{figure}

In the experiments, we launched five different adversarial attacks against six classes of networks (ref. Table ~\ref{Networks}) to highlight the transferability in a large domain. Concretely, for each two different classes of networks, namely the SN (Source Network) and the TN (Target Network), we analyzed the transferability property of adversarial attacks from the SN to the TN. According to Barni et al.~\cite{Barni2019OnForensics} and Papernot et al. \cite{PapernotTransferabilitySamples}, we split our experiments into three scenarios of transferability. These scenarios are related to the type of mismatch between the SN and the TN (i.e., half mismatch and complete mismatch). We provide in Figure ~\ref{transferability} the general scheme of the transferability property in computer networks.
\hfill
\begin{itemize}
    \item \textit{\textbf{Cross Training transferability}:} This scenario is a half mismatch where the SN and TN share the same network architecture. The both networks are trained on two different datasets, namely N-BaIoT and DGA datasets.
    \item \textit{\textbf{Cross Model transferability}:} The cross model transferability is also a half mismatch setting. In this scenario, we train the same dataset on two different architectures, namely \textit{SPRITZ1} and \textit{SPRITZ2}.
    \item \textit{\textbf{Cross Model and Training transferability}:} This scenario is a complete mismatch where the SN and TN have different architectures and also trained on different datasets.
\end{itemize}

The three scenarios resulted in a large set of experiments, and we discuss their outcomes in Section \ref{Results}. Table ~\ref{Cases} summarizes the scenarios presented above with their respective SN and TN. To simplify our experiments, we did not consider all possible scenarios. However, the reasonable number of experiments that we conducted is acceptable to conclude the transferability property in computer networks. 

\input{Tables/Cases}

%% file: Tables/Networks.tex
\begin{table}[!h]
\centering
\caption{General Description of the Considered Networks with their Respective Datasets. \label{Networks}}
\begin{tabular}{|c|c|c|c|}
\hline
\backslashbox{Networks}{Datasets} & \textbf{N-BaIoT} & \textbf{DGA}& \textbf{RIPE}\TBstrut\\ \hline

\textbf{SPRITZ1} & $N_{SPRITZ1}^{N-BaIoT}$ &  $N_{SPRITZ1}^{DGA}$ & $N_{SPRITZ1}^{RIPE}$\TBstrut\\\hline

\textbf{SPRITZ2} & $N_{SPRITZ2}^{N-BaIoT}$ & $N_{SPRITZ2}^{DGA}$ & $N_{SPRITZ2}^{RIPE}$\TBstrut\\\hline

\end{tabular}
\end{table}

%% file: Tables/Cases.tex
\begin{table}[!h]

\centering
\caption{Scenarios of the Considered Source and Target Networks.\label{Cases}}
\begin{threeparttable}
\begin{tabular}{|c|c|c|c|c|c|c|}
\hline
\backslashbox{\textbf{SN}}{\textbf{TN}} & \bm{$N_{1}$}  & \bm{$N_{2}$} & \bm{$N_{3}$} & \bm{$N_{4}$} & \bm{$N_{5}$} & \bm{$N_{6}$} \\ \hline

\bm{$N_{1}$} & \cellcolor{grey} & CT &  CM &   CMT & CT& \cellcolor{grey} \TBstrut\\\hline

\bm{$N_{2}$} &  CT &  \cellcolor{grey}& \cellcolor{grey} &   CM &\cellcolor{grey}&\cellcolor{grey}\TBstrut\\\hline

\bm{$N_{3}$} &  CM  & \cellcolor{grey} & \cellcolor{grey} &   CT &\cellcolor{grey}&\cellcolor{grey} \TBstrut\\\hline

\bm{$N_{4}$ }&  CMT  &  CM  &  CT & \cellcolor{grey} &\cellcolor{grey}&\cellcolor{grey}\TBstrut\\\hline

\bm{$N_{5}$} & \cellcolor{grey} &  \cellcolor{grey}  & \cellcolor{grey} & \cellcolor{grey} &\cellcolor{grey}&CM\TBstrut\\\hline

\bm{$N_{6}$} & \cellcolor{grey}  &  \cellcolor{grey}  & \cellcolor{grey} & \cellcolor{grey} &\cellcolor{grey}&CMT\TBstrut\\\hline

\end{tabular}

\begin{tablenotes}
      \small
      \item \small
    \bm{$N_{1}$}:$N_{SPRITZ1}^{N-BaIoT}$, \bm{$N_{2}$}:$N_{SPRITZ1}^{DGA}$, \bm{$N_{3}$}:$N_{SPRITZ2}^{N-BaIoT}$,  
\end{tablenotes}

\begin{tablenotes}
      \small
      \item \small \bm{$N_{4}$}:$N_{SPRITZ2}^{DGA}$ \bm{$N_{5}$}:$N_{SPRITZ1}^{RIPE}$,  \bm{$N_{6}$}:$N_{SPRITZ2}^{RIPE}$
    
\end{tablenotes}
\end{threeparttable}

\end{table}

%% file: Results_And_Discussion.tex
\section{Results and Discussion}
\label{Results}
In this section, we present the results of our experiments; then, we suggest two defense mechanisms against adversarial transferability. Afterward, we provide a discussion to evaluate the performance of the transferability property in computer networks.

\subsection{Experimental Results}

The test accuracy achieved by \textit{SPRITZ1} and \textit{SPRITZ2} on the test set N-BaIoT, DGA, and RIPE, detection task is: 99.89\% for $N_{SPRITZ1}^{N-BaIoT}$, $N_{SPRITZ2}^{N-BaIoT}$, $N_{SPRITZ2}^{DGA}$, and $N_{SPRITZ2}^{RIPE}$. We achieved the test accuracy 99.50\% for $N_{SPRITZ1}^{DGA}$, and 99.20\% for $N_{SPRITZ1}^{RIPE}$.\\
\indent For the I-FGSM attack, we fixed the number of steps to 10, and we considered the normalized strength factor $\varepsilon = 0.1$, 0.01, and 0.001, where the average PSNR is above 40 dB. Regarding the JSMA attack, the model parameters $\theta$ are set to 0.1 and 0.01. 
 
The results presented in the following correspond to the average ASR on SN, the average ASR on TN, the average PSNR on 250 samples, the average $L_1$ distortion, and the average maximum absolute distortion. During the experiments, we consider the adversarial attack transferability successful if the average ASR threshold is above 50\%. We highlight the transferable attacks in bold for each scenario.

\textbf{Cross Training transferability.} We illustrate and report in Table~\ref{CT_SPRITZ1_First} five case scenarios in the cross training transferability. For the \textit{SPRITZ1} network, and given both datasets interchangeably across the SN and the TN. Although the average ASR on the SN is successful (except for the JSMA attack that fails on the SN $N_{SPRITZ1}^{N-BaIoT}$ with the parameter $\theta = 0.01$), the transferability property is not satisfied on the TN. More specifically, for all the considered five adversarial attacks, the average ASR on the TN is almost null. Similarly, for the \textit{SPRITZ2} network, the adversarial attacks cannot be transferred between the SN and the TN. In particular, we remark that the average ASR on the TN remains almost null. Moreover, the JSMA attack fails on the SN $N_{SPRITZ2}^{N-BaIoT}$ and $N_{SPRITZ2}^{DGA}$ when considering the parameter $\theta = 0.01$. In summary, we observe that the adversarial attacks generated are not transferable from the SN to the TN, and in all the cases, the adversary cannot fool the TN.  With respect to the training dataset, it is worth mentioning that, for a given computer network, where the SN and TN share the same architecture and are trained on two different datasets, the cross training transferability is asymmetric. 
In other words, when the datasets are mismatched but the architectures are matched, the I-FGSM, JSMA, LBFGS, PGD, and DeepFool attacks are less transferable in the cross-training transferability. One plausible explanation is that while the attacks only affect a small percentage of the samples, the attacked model tends to overfit more. It's also worth noting that the transferability of training datasets for a specific task isn't symmetric. This implies that, mostly in computer networks, the baseline dataset may influence the features learned by the network in some way and to some amount.
We considered another dataset, RIPE Atlas, as a TN (\textit{SPRITZ1}) for a few experiments to better evaluate the transferability performance. When using RIPE Atlas dataset, we have found that attacking on SN is still not enough to deceive TN.

\textbf{Cross Model transferability.} We consider five case scenarios in the cross model transferability, and we describe the experimental results in Table ~\ref{CM_BaIoT_First}. When training the same datasets on two different networks (\textit{SPRITZ1} and \textit{SPRITZ2}), we can see that the average ASR on the SN is successful (except for the JSMA attack that fails on the SN $N_{SPRITZ1}^{N-BaIoT}$ with the parameter $\theta = 0.01$). Moreover, we remark that some adversarial attacks exhibit a transferability between the SN and the TN, such as the JSMA attack, the I-FGSM attack, and the LBFGS attack. For the N-BaIoT datasets, the JSMA attack enables a strong transferability from the SN to the TN when considering the SNs $N_{SPRITZ1}^{N-BaIoT}$ and  $N_{SPRITZ2}^{N-BaIoT}$ with the parameter $\theta = 0.1$ and $\theta = 0.01$ respectively. The experimental results show that the average ASR on the TN is $1$ for $N_{SPRITZ1}^{N-BaIoT}$ as a SN and $0.9960$ for $N_{SPRITZ2}^{N-BaIoT}$ as a SN. Regarding the DGA datasets, the average ASR on the SN is successful (except for the JSMA attack that fails on the SN $N_{SPRITZ2}^{DGA}$ with the parameter $\theta = 0.01$), and we achieve the transferability for both the I-FGSM attack and the LBFGS attack. Considering the $N_{SPRITZ2}^{DGA}$ as a SN, the transferability happens for the I-FGSM attack with both parameters $\varepsilon = 0.01$ and $\varepsilon = 0.001$, and their average ASR on the TN is $1$. Additionally, the LBFGS attack with the default parameter is transferable given the $N_{SPRITZ2}^{DGA}$ as a SN, and the average ASR on the TN is $0.63$. Differently from the previous scenario, the cross model transferability demonstrates five different cases in computer networks where the adversary can perfectly fool the TN. Therefore, the absence of transferability in the JSMA, IFGSM, and LBFGS attacks scenarios in cross-modal transferability is considerably more significant. As a result, we may expect shallow and deep architectures to obtain similar peculiar features in some scenarios, thus enhancing the attack transferability. For additional tests, we used the RIPE dataset in this setting while TN was trained on it. We noticed that non-transferability exists in this scenario as well, when RIPE dataset is employed.
\input{Tables/CT_SPRITZ1_First}
\input{Tables/CM_BaIoT_First}

\input{Tables/CMT}

\begin{figure}[!h]
\begin{center}
\includegraphics[width=0.48\textwidth]{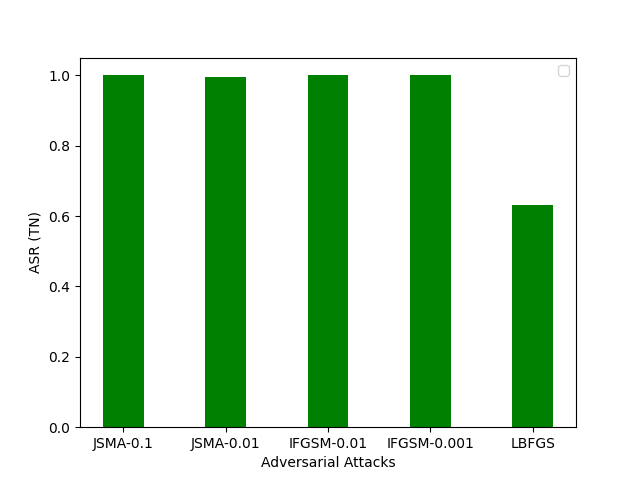}
\caption{Summary of transferability cases where adversarial attacks transfer between SN to TN.}
\label{VisualRep}
\end{center}
\end{figure}

\textbf{Cross Model and Training transferability.} We consider three case scenarios, and present in Table ~\ref{CMT} our evaluation on two different networks (\textit{SPRITZ1} and \textit{SPRITZ2}) that are trained on three different datasets (N-BaIoT, DGA, and RIPE Atlas). As it can be seen, the average ASR on the SN is successful (except for the JSMA attack that fails on the SN $N_{SPRITZ1}^{N-BaIoT}$ with the parameter $\theta = 0.01$, and the SN $N_{SPRITZ2}^{DGA}$ with the parameters $\theta = 0.01$). We notice that none of the five adversarial attacks performed in our experiments are transferable from the SN to the TN, and the average ASR on the TN is almost null. Therefore, the feasibility of the cross model and training transferability in computer networks is unlikely hard to happen. Table~\ref{CMT} shows that, in this scenario, the attack success rate decreases, as we have a full mismatch in networks and datasets. 
For additional experiments, we used the RIPE as a mismatch dataset for TN. In a complete mismatch setting, we conclude that SN and TN are non-transferable. 
Figure~\ref{VisualRep} shows a visual representation of adversarial attacks that transfer from SN to TN.

\subsection{Anti-Exploratory Evasion Defense mechanisms}
The outcomes of our study reveal that the adversarial samples are often non-transferable, contrary to what we observe in standard classification tasks, such as multimedia forensics and similar applications. This finding is important because we can leverage the limitation of transferability to hinder the attack. For instance, an LK scenario could be imposed somehow, as with standard ML techniques, to prevent adversarial examples. However, we identified various adversarial attacks in different scenarios that enable the transferability property, meaning that the adversarial attacks can entirely mislead the TN.
In this case, we suggest two current defense mechanisms that prevent adversarial transferability. The first strategy consists of fine-tuning the classifier using the Most Powerful Attacks (MPAs) whenever the transferability happens for a given adversarial attack~\cite{Barni2017MPA}. Consequently, we improve the security of the TN against different adversarial attacks only for the attacks that transfer between the SN and TN (refer to Table VI). We apply the fine-tuning technique on the TN, and we report our results in Table~\ref{Defense_MPA}.
Another strategy to prevent adversarial transferability relies on using the LSTM architecture as the TN rather than the CNN, namely as mismatch architecture. Therefore, the TN is different from the classification structure of the original CNN. In this strategy, the attacker has less attack information than the previous system with no knowledge of the TN architecture. We verify this approach in a Cross-Model setting by considering two adversarial attacks enabling the transferability: the JSMA attack with a parameter $\theta$ = 0.01 and the LBFGS attack with a default parameter. Then, we changed the TN architecture from CNN to LSTM. In this scenario, the ASR on the TN significantly decreased in the JSMA attack from 0.9960 to 0.2360 and in the LBFGS attack from 0.6300 to 0.0010. The outcomes of this technique demonstrate that changing the architecture (i.e., SN (CNN) and TN (LSTM)) can significantly prevent adversarial attacks transferability and hence improve the security of the TN. Consequently, the architectural mismatch between the original CNN and the new detector TN is sufficient to prohibit transferability, which may be considered as a shielding strategy.

\input{Tables/DefenseMPA}
\subsection{Discussion}
\input{Tables/executionTime}
Combining all the experimental results, we demonstrate that among five different adversarial attacks, three of them are transferable from the SN to the TN, which are: the JSMA attack, the I-FGSM attack, and the LBFGS attack. Therefore, given a CNN-based model in computer network, specific cases of generated adversarial examples can perfectly fool another network model even with different architectures. In particular, we found that both models in the SN and TN enable the same misclassification; hence, the transferability property. To that end, the adversary can leverage a surrogate model with different architectures and classes of algorithms to approximate the adversarial example for the TN. We note that in the cross model transferability, the features learned by the network might influence somehow to some degree by the underlying dataset, and this aspect requires additional research.

Moreover, we performed multiple adversarial attacks on the SN for 250 manipulated examples to evaluate their computational cost. On the other hand, after detecting adversarial attacks on SN, we remark that the processing testing time for TN is relatively fast. In Table~\ref{Exe_Time}, we report our results on the execution running attack time on SN in seconds. This study, as we expected, will also address the issue of usability in botnet attacks. To elaborate more, the authors in~\cite{Karaca2021} suggested a solution for botnet attack detection in the IoT environment using CNN. As a result, most ML and DL methods are subject to various adversarial attacks if we have a classification algorithm for botnet detection. Additionally, the experts in the research area demonstrate that an attack in the SN might not be enough to deceive the TN, based on our studies and other research in computer networks.

%% file: Tables/CT_SPRITZ1_First.tex
\begin{table*}[!h]
\centering
\caption{Experimental Results for Cross Training Transferability on SPRITZ1 and SPRITZ2 Architecture\\ (N-BaIoT and DGA datasets are used interchangeably in the SN and the TN, while RIPE Atlas dataset is used only in the TN)}. \label{CT_SPRITZ1_First}
\resizebox{2\columnwidth}{!}{
\begin{tabular}{|c|c|c|c|c|c|c|c|c|}
\hline
\textbf{SN} & \textbf{TN} & \textbf{Accuracy} & \textbf{Attack Type} & \textbf{PSNR} & \textbf{$L_1$ dist} & \textbf{Max. dist} & \textbf{ASR} & \textbf{ASR} \TBstrut\\
& & & & & & & \textbf{(SN)} & \textbf{(TN)}  \\ \hline
 \rowcolor{grey}
$N_{SPRITZ1}^{N-BaIoT}$ & $N_{SPRITZ1}^{DGA}$ & SN = 98.90\% , TN = 100\% & I-FGSM, $\varepsilon$ = 0.1 & 18.65 & 22.07 & 65.12 & 1.0000 & 0.0480 \TBstrut\\ \hline

$N_{SPRITZ1}^{N-BaIoT}$ & $N_{SPRITZ1}^{DGA}$ & SN = 98.90\% , TN = 100\% & I-FGSM, $\varepsilon$ = 0.01 & 18.93 & 21.26 & 62.71 & 1.0000 & 0.0000 \TBstrut\\ \hline
 \rowcolor{grey}
$N_{SPRITZ1}^{N-BaIoT}$ & $N_{SPRITZ1}^{DGA}$ & SN = 98.90\% , TN = 100\% & I-FGSM, $\varepsilon$ = 0.001 & 19.06 & 20.94 & 61.76 & 1.0000 & 0.0000 \TBstrut\\ \hline

$N_{SPRITZ1}^{N-BaIoT}$ & $N_{SPRITZ1}^{DGA}$ & SN = 98.90\% , TN = 100\% & JSMA, $\theta$ = 0.1 & 18.49 & 7.29 & 178.50 & 1.0000 & 0.1100 \TBstrut\\ \hline
 \rowcolor{grey}
$N_{SPRITZ1}^{N-BaIoT}$ & $N_{SPRITZ1}^{DGA}$ & SN = 98.90\% , TN = 100\% & JSMA, $\theta$ = 0.01 & -- & -- & -- & FAIL\textsuperscript{*} & -- \TBstrut\\ \hline

$N_{SPRITZ1}^{N-BaIoT}$ & $N_{SPRITZ1}^{DGA}$ & SN = 98.90\% , TN = 100\% & LBFGS, default parameter & 22.57 & 9.98 & 106.35 & 1.0000 & 0.0000 \TBstrut\\ \hline
 \rowcolor{grey}
$N_{SPRITZ1}^{N-BaIoT}$ & $N_{SPRITZ1}^{DGA}$ & SN = 98.90\% , TN = 100\% & PGD, default parameter & 19.59 & 19.71 & 54.00 & 0.9800 & 0.0020 \TBstrut\\ \hline

$N_{SPRITZ1}^{N-BaIoT}$ & $N_{SPRITZ1}^{DGA}$ & SN = 98.90\% , TN = 100\% & DeepFool, default parameter & 20.90 & 12.11 & 142.85 & 0.9500 & 0.0400 \TBstrut\\ \hline

 \rowcolor{grey}
$N_{SPRITZ1}^{DGA}$ & $N_{SPRITZ1}^{N-BaIoT}$ & SN = 100\% , TN = 98.90\% & I-FGSM, $\varepsilon$ = 0.1 & 31.4618 & 4.81 & 17.20 & 1.0000 & 0.0050 \TBstrut\\ \hline

$N_{SPRITZ1}^{DGA}$ & $N_{SPRITZ1}^{N-BaIoT}$ & SN = 100\% , TN = 98.90\% & I-FGSM, $\varepsilon$ = 0.01 & 32.51 & 4.21 & 15.92 & 1.0000 & 0.0000 \TBstrut\\ \hline
 \rowcolor{grey}
$N_{SPRITZ1}^{DGA}$ & $N_{SPRITZ1}^{N-BaIoT}$ & SN = 100\% , TN = 98.90\% & I-FGSM, $\varepsilon$ = 0.001 & 32.96 & 4.00 & 15.03 & 1.0000 & 0.0000 \TBstrut\\ \hline

$N_{SPRITZ1}^{DGA}$ & $N_{SPRITZ1}^{N-BaIoT}$ & SN = 100\% , TN = 98.90\% & JSMA, $\theta$ = 0.1 & 31.23 & 1.47 & 62.50 & 1.0000 & 0.2000 \TBstrut\\ \hline
 \rowcolor{grey}
$N_{SPRITZ1}^{DGA}$ & $N_{SPRITZ1}^{N-BaIoT}$ & SN = 100\% , TN = 98.90\% & JSMA, $\theta$ = 0.01 & 35.95 & 1.077 & 17.85 & 1.0000 & 0.0010 \TBstrut\\ \hline

$N_{SPRITZ1}^{DGA}$ & $N_{SPRITZ1}^{N-BaIoT}$ & SN = 100\% , TN = 98.90\% & LBFGS, default parameter & 37.19 & 1.41 & 23.38 & 1.0000 & 0.0000 \TBstrut\\ \hline
 \rowcolor{grey}
$N_{SPRITZ1}^{DGA}$ & $N_{SPRITZ1}^{N-BaIoT}$ & SN = 100\% , TN = 98.90\% & PGD, default parameter &  33.75 & 3.58 & 13.93 & 1.0000 & 0.0000\TBstrut\\ \hline

$N_{SPRITZ1}^{DGA}$ & $N_{SPRITZ1}^{N-BaIoT}$ & SN = 100\% , TN = 98.90\% & DeepFool, default parameter &  36.98 & 2.20 & 38.08 & 1.0000 & 0.0300 \TBstrut\\ \hline

 \rowcolor{grey}
$N_{SPRITZ2}^{N-BaIoT}$ & $N_{SPRITZ2}^{DGA}$ & SN = 98.90\% , TN = 100\% & I-FGSM, $\varepsilon$ = 0.1 & 17.98 & 22.67 & 58.06  & 1.0000 & 0.0360  \TBstrut\\ \hline

$N_{SPRITZ2}^{N-BaIoT}$ & $N_{SPRITZ2}^{DGA}$ & SN = 98.90\% , TN = 100\% & I-FGSM, $\varepsilon$ = 0.01 & 18.64 & 20.96 & 53.47  & 1.0000 & 0.0000 \TBstrut\\ \hline
 \rowcolor{grey}
$N_{SPRITZ2}^{N-BaIoT}$ & $N_{SPRITZ2}^{DGA}$ & SN = 98.90\% , TN = 100\% & I-FGSM, $\varepsilon$ = 0.001 & 18.71 & 20.82 & 53.16  & 1.0000 & 0.0000 \TBstrut\\ \hline

$N_{SPRITZ2}^{N-BaIoT}$ & $N_{SPRITZ2}^{DGA}$ & SN = 98.90\% , TN = 100\% & JSMA, $\theta$ = 0.1 & 22.23 & 2.91  & 178.08 & 1.0000 & 0.0061 \TBstrut\\ \hline
 \rowcolor{grey}
$N_{SPRITZ2}^{N-BaIoT}$ & $N_{SPRITZ2}^{DGA}$ & SN = 98.90\% , TN = 100\% & JSMA, $\theta$ = 0.01 & -- & -- & -- & FAIL\textsuperscript{*} & -- \TBstrut\\ \hline

$N_{SPRITZ2}^{N-BaIoT}$ & $N_{SPRITZ2}^{DGA}$ & SN = 98.90\% , TN = 100\% & LBFGS, default parameter & 24.62 & 4.67  & 194.25 & 1.0000 & 0.0000 \TBstrut\\ \hline
 \rowcolor{grey}
$N_{SPRITZ2}^{N-BaIoT}$ & $N_{SPRITZ2}^{DGA}$ & SN = 98.90\% , TN = 100\% & PGD, default parameter & 18.75 & 20.66 & 53.07  & 0.9916 & 0.1812 \TBstrut\\ \hline

$N_{SPRITZ2}^{N-BaIoT}$ & $N_{SPRITZ2}^{DGA}$ & SN = 98.90\% , TN = 100\% & DeepFool, default parameter & 23.86 & 4.59  & 177.38 & 1.0000 & 0.0000 \TBstrut\\ \hline
 \rowcolor{grey}
$N_{SPRITZ2}^{DGA}$ & $N_{SPRITZ2}^{N-BaIoT}$ & SN = 100\% , TN = 98.90\% & I-FGSM, $\varepsilon$ = 0.1 & 28.48 & 6.03 & 25.15 & 0.9800 & 0.0150\TBstrut\\ \hline

$N_{SPRITZ2}^{DGA}$ & $N_{SPRITZ2}^{N-BaIoT}$ & SN = 100\% , TN = 98.90\% & I-FGSM, $\varepsilon$ = 0.01 &  28.61 & 5.99 & 23.34 & 1.0000 & 0.0000 \TBstrut\\ \hline
 \rowcolor{grey}
$N_{SPRITZ2}^{DGA}$ & $N_{SPRITZ2}^{N-BaIoT}$ & SN = 100\% , TN = 98.90\% & I-FGSM, $\varepsilon$ = 0.001 &  28.92 & 5.76 & 22.64 & 1.0000 & 0.0000\TBstrut\\ \hline

$N_{SPRITZ2}^{DGA}$ & $N_{SPRITZ2}^{N-BaIoT}$ & SN = 100\% , TN = 98.90\% & JSMA, $\theta$ = 0.1 & 31.17 & 1.28 & 56.56 & 1.0000 & 0.1000 \TBstrut\\ \hline
 \rowcolor{grey}
$N_{SPRITZ2}^{DGA}$ & $N_{SPRITZ2}^{N-BaIoT}$ & SN = 100\% , TN = 98.90\% & JSMA, $\theta$ = 0.01 & -- & -- & -- & FAIL\textsuperscript{*} & -- \TBstrut\\ \hline

$N_{SPRITZ2}^{DGA}$ & $N_{SPRITZ2}^{N-BaIoT}$ & SN = 100\% , TN = 98.90\% & LBFGS, default parameter & 34.22 & 1.40 & 33.93 & 1.0000 & 0.0000 \TBstrut\\ \hline
 \rowcolor{grey}
$N_{SPRITZ2}^{DGA}$ & $N_{SPRITZ2}^{N-BaIoT}$ & SN = 100\% , TN = 98.90\% & PGD, default parameter & 29.24 & 5.49 & 22.24 & 1.0000 & 0.0000 \TBstrut\\ \hline

$N_{SPRITZ2}^{DGA}$ & $N_{SPRITZ2}^{N-BaIoT}$ & SN = 100\% , TN = 98.90\% & DeepFool, default parameter & 31.02 & 2.25 & 53.42 & 0.9812 & 0.1200  \TBstrut\\ \hline

 \rowcolor{grey}
$N_{SPRITZ1}^{N-BaIoT}$ & $N_{SPRITZ1}^{RIPE}$ & SN = 98.90\% , TN = 99.20\% & I-FGSM, $\varepsilon$ = 0.1 & 18.65 & 22.07 & 65.12 & 1.0000 & 0.0200 \TBstrut\\ \hline

$N_{SPRITZ1}^{N-BaIoT}$ & $N_{SPRITZ1}^{RIPE}$ & SN = 98.90\% , TN = 99.20\% & JSMA, $\theta$ = 0.1 & 18.49 & 7.29 & 178.50 & 1.0000 & 0.0000 \TBstrut\\ \hline

$N_{SPRITZ1}^{N-BaIoT}$ & $N_{SPRITZ1}^{RIPE}$ & SN = 98.90\% , TN = 99.20\% & LBFGS, default parameter & 22.57 & 9.98 & 106.35 & 1.0000 & 0.0010 \TBstrut\\ \hline

 \rowcolor{grey}
$N_{SPRITZ1}^{N-BaIoT}$ & $N_{SPRITZ1}^{RIPE}$ & SN = 98.90\% , TN = 99.20\% & PGD, default parameter & 19.59 & 19.71 & 54.00 & 0.9800 & 0.0000 \TBstrut\\ \hline

$N_{SPRITZ1}^{DGA}$ & $N_{SPRITZ1}^{RIPE}$ & SN = 98.90\% , TN = 99.20\% & DeepFool, default parameter &  36.98 & 2.20 & 38.08 & 1.0000 & 0.0000 \TBstrut\\ \hline

\end{tabular}}
\\FAIL\textsuperscript{*}: In this case, the attack on the source network fails and we cannot transfer it to the target network. Therefore, the numerical values of the average PSNR, the average $L_1$ dist, and the average maximum distortion are not possible.  
\end{table*}

%% file: Tables/CM_BaIoT_First.tex
\begin{table*}[!h]
\centering
\caption{Experimental Results for Cross Model Transferability on N-BaIoT, DGA, and RIPE Atlas datasets\\ (SPRITZ1 and SPRITZ2 are used interchangeably in the SN and the TN). \label{CM_BaIoT_First}}
\resizebox{2\columnwidth}{!}{
\begin{tabular}{|c|c|c|c|c|c|c|c|c|}
\hline
\textbf{SN} & \textbf{TN} & \textbf{Accuracy} & \textbf{Attack Type} & \textbf{PSNR} & \textbf{$L_1$ dist} & \textbf{Max. dist} & \textbf{ASR} & \textbf{ASR} \TBstrut\\
& & & & & & & \textbf{(SN)} & \textbf{(TN)}  \\ \hline
 \rowcolor{grey}
$N_{SPRITZ1}^{N-BaIoT}$ & $N_{SPRITZ2}^{N-BaIoT}$ & SN = 98.90\% , TN = 98.80\% & I-FGSM, $\varepsilon$ = 0.1 & 18.65 & 22.07 & 64.78  & 1.0000 & 0.0030 \TBstrut\\ \hline

$N_{SPRITZ1}^{N-BaIoT}$ & $N_{SPRITZ2}^{N-BaIoT}$ & SN = 98.90\% , TN = 98.80\% & I-FGSM, $\varepsilon$ = 0.01 & 18.93 & 21.26 & 62.71  & 1.0000 & 0.0000 \TBstrut\\ \hline
 \rowcolor{grey}
$N_{SPRITZ1}^{N-BaIoT}$ & $N_{SPRITZ2}^{N-BaIoT}$ & SN = 98.90\% , TN = 98.80\% & I-FGSM, $\varepsilon$ = 0.001 &  19.04 & 20.94 & 61.76  & 1.0000 & 0.0000 \TBstrut\\ \hline

$N_{SPRITZ1}^{N-BaIoT}$ & $N_{SPRITZ2}^{N-BaIoT}$ & SN = 98.90\% , TN = 98.80\% & \textbf{JSMA, $\theta$ = 0.1} & \textbf{18.49} & \textbf{7.29}  & \textbf{170.50} & \textbf{1.0000} & \textbf{1.0000} \TBstrut\\ \hline
 \rowcolor{grey}
$N_{SPRITZ1}^{N-BaIoT}$ & $N_{SPRITZ2}^{N-BaIoT}$ & SN = 98.90\% , TN = 98.80\% & JSMA, $\theta$ = 0.01 & --     & --     & --      & FAIL\textsuperscript{*}   & --      \TBstrut\\ \hline

$N_{SPRITZ1}^{N-BaIoT}$ & $N_{SPRITZ2}^{N-BaIoT}$ & SN = 98.90\% , TN = 98.80\% & LBFGS, default parameter & 22.57 & 9.98  & 106.42 & 0.9960 & 0.0040 \TBstrut\\ \hline
 \rowcolor{grey}
$N_{SPRITZ1}^{N-BaIoT}$ & $N_{SPRITZ2}^{N-BaIoT}$ & SN = 98.90\% , TN = 98.80\% & PGD, default parameter & 19.59 & 19.71 & 54.00  & 1.0000 & 0.0104 \TBstrut\\ \hline

$N_{SPRITZ1}^{N-BaIoT}$ & $N_{SPRITZ2}^{N-BaIoT}$ & SN = 98.90\% , TN = 98.80\% & Deepfool, default parameter & 20.86 & 12.15 & 143.31 & 0.9190 & 0.1336 \TBstrut\\ \hline

 \rowcolor{grey}
$N_{SPRITZ2}^{N-BaIoT}$ & $N_{SPRITZ1}^{N-BaIoT}$ & SN = 98.90\% , TN = 98.80\% & I-FGSM, $\varepsilon$ = 0.1 & 17.98 & 22.67 & 58.06 & 0.9913 & 0.0200 \TBstrut\\ \hline

$N_{SPRITZ2}^{N-BaIoT}$ & $N_{SPRITZ1}^{N-BaIoT}$ & SN = 98.90\% , TN = 98.80\% & I-FGSM, $\varepsilon$ = 0.01 & 18.65 & 20.96 & 53.47 & 1.0000 & 0.0000 \TBstrut\\ \hline
 \rowcolor{grey}
$N_{SPRITZ2}^{N-BaIoT}$ & $N_{SPRITZ1}^{N-BaIoT}$ & SN = 98.90\% , TN = 98.80\% & I-FGSM, $\varepsilon$ = 0.001 & 18.71 & 20.82 & 53.16 & 1.0000 & 0.0000\TBstrut\\ \hline

$N_{SPRITZ2}^{N-BaIoT}$ & $N_{SPRITZ1}^{N-BaIoT}$ & SN = 98.90\% , TN = 98.80\% & JSMA, $\theta$ = 0.1 & 22.23 & 2.91 & 178.08 & 1.0000 & 0.0000 \TBstrut\\ \hline
 \rowcolor{grey}
$N_{SPRITZ2}^{N-BaIoT}$ & $N_{SPRITZ1}^{N-BaIoT}$ & SN = 98.90\% , TN = 98.80\% & \textbf{JSMA, $\theta$ = 0.01} &  \textbf{40.93} & \textbf{0.30} & \textbf{17.84} & \textbf{1.0000} & \textbf{0.9960} \TBstrut\\ \hline

$N_{SPRITZ2}^{N-BaIoT}$ & $N_{SPRITZ1}^{N-BaIoT}$ & SN = 98.90\% , TN = 98.80\% & LBFGS, default parameter &  24.62 & 4.67 & 114.36 & 1.0000 & 0.0000\TBstrut\\ \hline
 \rowcolor{grey}
$N_{SPRITZ2}^{N-BaIoT}$ & $N_{SPRITZ1}^{N-BaIoT}$ & SN = 98.90\% , TN = 98.80\% & PGD, default parameter & 18.75 & 20.66 & 53.07 & 1.0000 & 0.0000 \TBstrut\\ \hline

$N_{SPRITZ2}^{N-BaIoT}$ & $N_{SPRITZ1}^{N-BaIoT}$ & SN = 98.90\% , TN = 98.80\% & DeepFool, default parameter & 23.86 & 4.59 & 177.38 & 1.0000 & 0.0000 \TBstrut\\ \hline

 \rowcolor{grey}
$N_{SPRITZ1}^{DGA}$ & $N_{SPRITZ2}^{DGA}$ & SN = 99.60\% , TN = 99.60\% & I-FGSM, $\varepsilon$ = 0.1 & 31.46 & 4.81 & 17.20 & 1.0000 & 0.0100 \TBstrut\\ \hline

$N_{SPRITZ1}^{DGA}$ & $N_{SPRITZ2}^{DGA}$ & SN = 99.60\% , TN = 99.60\% & I-FGSM, $\varepsilon$ = 0.01 & 32.51 & 4.21 & 15.92 & 1.0000 & 0.0000 \TBstrut\\ \hline
 \rowcolor{grey}
$N_{SPRITZ1}^{DGA}$ & $N_{SPRITZ2}^{DGA}$ & SN = 99.60\% , TN = 99.60\% & I-FGSM, $\varepsilon$ = 0.001 &  32.96 & 4.00 & 15.03 & 1.0000 & 0.0000 \TBstrut\\ \hline

$N_{SPRITZ1}^{DGA}$ & $N_{SPRITZ2}^{DGA}$ & SN = 99.60\% , TN = 99.60\% & JSMA, $\theta$ = 0.1 & 31.23 & 1.47 & 62.50 & 1.0000 & 0.5020 \TBstrut\\ \hline
 \rowcolor{grey}
$N_{SPRITZ1}^{DGA}$ & $N_{SPRITZ2}^{DGA}$ & SN = 99.60\% , TN = 99.60\% & JSMA, $\theta$ = 0.01 & 35.95 & 1.07 & 17.85 & 1.0000 & 0.0900 \TBstrut\\ \hline

$N_{SPRITZ1}^{DGA}$ & $N_{SPRITZ2}^{DGA}$ & SN = 99.60\% , TN = 99.60\% & LBFGS, default parameter & 37.19 & 1.47  & 23.20 & 0.9930 & 0.0400 \TBstrut\\ \hline
 \rowcolor{grey}
$N_{SPRITZ1}^{DGA}$ & $N_{SPRITZ2}^{DGA}$ & SN = 99.60\% , TN = 99.60\% & PGD, default parameter & 33.75 & 3.58 & 13.93 & 1.0000 & 0.0000 \TBstrut\\ \hline

$N_{SPRITZ1}^{DGA}$ & $N_{SPRITZ2}^{DGA}$ & SN = 99.60\% , TN = 99.60\% & DeepFool, default parameter & 34.11 & 2.20 & 38.21 & 1.0000 & 0.0000 \TBstrut\\ \hline

 \rowcolor{grey}
$N_{SPRITZ2}^{DGA}$ & $N_{SPRITZ1}^{DGA}$ & SN = 99.60\% , TN = 99.60\% & I-FGSM, $\varepsilon$ = 0.1 &  28.48 & 6.04 & 24.15 & 1.0000 & 0.4100 \TBstrut\\ \hline

$N_{SPRITZ2}^{DGA}$ & $N_{SPRITZ1}^{DGA}$ & SN = 99.60\% , TN = 99.60\% & \textbf{I-FGSM, $\varepsilon$ = 0.01} & \textbf{28.62} & \textbf{5.99} & \textbf{23.25} & \textbf{1.0000} & \textbf{1.0000} \TBstrut\\ \hline
 \rowcolor{grey}
$N_{SPRITZ2}^{DGA}$ & $N_{SPRITZ1}^{DGA}$ & SN = 99.60\% , TN = 99.60\% & \textbf{I-FGSM, $\varepsilon$ = 0.001} & \textbf{28.92} & \textbf{5.76} & \textbf{22.64} & \textbf{1.0000} & \textbf{1.0000} \TBstrut\\ \hline

$N_{SPRITZ2}^{DGA}$ & $N_{SPRITZ1}^{DGA}$ & SN = 99.60\% , TN = 99.60\% & JSMA, $\theta$ = 0.1 & 31.17 & 1.28 & 56.56 & 1.0000 & 0.0000  \TBstrut\\ \hline
 \rowcolor{grey}
$N_{SPRITZ2}^{DGA}$ & $N_{SPRITZ1}^{DGA}$ & SN = 99.60\% , TN = 99.60\% & JSMA, $\theta$ = 0.01 & --     & --     & --      & FAIL\textsuperscript{*}   & --  \TBstrut\\ \hline

$N_{SPRITZ2}^{DGA}$ & $N_{SPRITZ1}^{DGA}$ & SN = 99.60\% , TN = 99.60\% & \textbf{LBFGS, default parameter} &  \textbf{34.22} & \textbf{1.48} & \textbf{33.92} & \textbf{0.9930} & \textbf{0.6300} \TBstrut\\ \hline
 \rowcolor{grey}
$N_{SPRITZ2}^{DGA}$ & $N_{SPRITZ1}^{DGA}$ & SN = 99.60\% , TN = 99.60\% & PGD, default parameter & 29.24 & 5.49 & 22.24 & 1.0000 & 0.3200  \TBstrut\\ \hline

$N_{SPRITZ2}^{DGA}$ & $N_{SPRITZ1}^{DGA}$ & SN = 99.60\% , TN = 99.60\% & DeepFool, default parameter & 31.02 & 2.25 & 53.42 & 1.0000 & 0.2329   \TBstrut\\ \hline

\rowcolor{grey}
$N_{SPRITZ1}^{RIPE}$ & $N_{SPRITZ2}^{RIPE}$ & SN = 98.90\% , TN = 99.80\% & I-FGSM, $\varepsilon$ = 0.1 & 18.65 & 22.07 & 64.78  & 1.0000 & 0.0000 \TBstrut\\ \hline

$N_{SPRITZ1}^{RIPE}$ & $N_{SPRITZ2}^{RIPE}$ & SN = 98.90\% , TN = 99.80\% & JSMA, $\theta$ = 0.1 & 18.49 & 7.29  & 170.50 & 1.0000 & 0.0140 \TBstrut\\ \hline

\rowcolor{grey}
$N_{SPRITZ1}^{RIPE}$ & $N_{SPRITZ2}^{RIPE}$ & SN = 98.90\% , TN = 99.80\% & LBFGS, default parameter & 22.57 & 9.98  & 106.42 & 0.9960 & 0.0100 \TBstrut\\ \hline

$N_{SPRITZ1}^{RIPE}$ & $N_{SPRITZ2}^{RIPE}$ & SN = 99.60\% , TN = 99.80\% & PGD, default parameter & 33.75 & 3.58 & 13.93 & 1.0000 & 0.0020 \TBstrut\\ \hline

\rowcolor{grey}
$N_{SPRITZ1}^{RIPE}$ & $N_{SPRITZ2}^{RIPE}$ & SN = 99.60\% , TN = 99.80\% & DeepFool, default parameter & 34.11 & 2.20 & 38.21 & 1.0000 & 0.0000 \TBstrut\\ \hline

\end{tabular}}
\\FAIL\textsuperscript{*}: In this case, the attack on the source network fails and we cannot transfer it to the target network. Therefore, the numerical values of the average PSNR, the average $L_1$ dist, and the average maximum distortion are not possible.  
\end{table*}

%% file: Tables/CMT.tex
\begin{table*}[!h]
\centering
\caption{Experimental results for Cross Model and Training Transferability (The architectures and datasets\\ are used interchangeably in the SN an the TN, except for RIPE Atlas dataset which is used only in the TN)
\label{CMT}}
\resizebox{2\columnwidth}{!}{
\begin{tabular}{|c|c|c|c|c|c|c|c|c|}
\hline
\textbf{SN} & \textbf{TN} & \textbf{Accuracy} & \textbf{Attack Type} & \textbf{PSNR} & \textbf{$L_1$ dist} & \textbf{Max. dist} & \textbf{ASR} & \textbf{ASR} \TBstrut\\
& & & & & & & \textbf{(SN)} & \textbf{(TN)}  \\ \hline
 \rowcolor{grey}
$N_{SPRITZ1}^{N-BaIoT}$ & $N_{SPRITZ2}^{DGA}$ & SN = 98.80\% , TN = 100\% & I-FGSM, $\varepsilon$ = 0.1 & 18.65 & 22.07 & 64.78  & 1.0000 & 0.0030 \TBstrut\\ \hline

$N_{SPRITZ1}^{N-BaIoT}$ & $N_{SPRITZ2}^{DGA}$ & SN = 98.80\% , TN = 100\% & I-FGSM, $\varepsilon$ = 0.01 & 18.93 & 21.26 & 62.71  & 1.0000 & 0.0000 \TBstrut\\ \hline
 \rowcolor{grey}
$N_{SPRITZ1}^{N-BaIoT}$ & $N_{SPRITZ2}^{DGA}$ & SN = 98.80\% , TN = 100\% & I-FGSM, $\varepsilon$ = 0.001 & 19.06 & 20.93 & 61.72  & 1.0000 & 0.0000 \TBstrut\\ \hline

$N_{SPRITZ1}^{N-BaIoT}$ & $N_{SPRITZ2}^{DGA}$ & SN = 98.80\% , TN = 100\% & JSMA, $\theta$ = 0.1 & 18.49 & 7.29  & 178.50 & 1.0000 & 0.0000 \TBstrut\\ \hline
 \rowcolor{grey}
$N_{SPRITZ1}^{N-BaIoT}$ & $N_{SPRITZ2}^{DGA}$ & SN = 98.80\% , TN = 100\% & JSMA, $\theta$ = 0.01 & --     & --     & --      & FAIL\textsuperscript{*}   & --      \TBstrut\\ \hline

$N_{SPRITZ1}^{N-BaIoT}$ & $N_{SPRITZ2}^{DGA}$ & SN = 98.80\% , TN = 100\% & LBFGS, default parameter & 22.57 & 10.46 & 110.67 & 0.9910 & 0.0000 \TBstrut\\ \hline
 \rowcolor{grey}
$N_{SPRITZ1}^{N-BaIoT}$ & $N_{SPRITZ2}^{DGA}$ & SN = 98.80\% , TN = 100\% & PGD, default parameter & 19.59 & 19.71 & 54.00  & 1.0000 & 0.0000 \TBstrut\\ \hline

$N_{SPRITZ1}^{N-BaIoT}$ & $N_{SPRITZ2}^{DGA}$ & SN = 98.80\% , TN = 100\% & DeepFool, default parameter & 20.87 & 12.14 & 143.19 & 0.9676 & 0.0324 \TBstrut\\ \hline

 \rowcolor{grey}
$N_{SPRITZ2}^{DGA}$ & $N_{SPRITZ1}^{N-BaIoT}$ & SN = 99.60\% , TN = 99.60\% & I-FGSM, $\varepsilon$ = 0.1 & 28.49 & 6.03 & 24.15 & 1.0000 & 0.0400 \TBstrut\\ \hline

$N_{SPRITZ2}^{DGA}$ & $N_{SPRITZ1}^{N-BaIoT}$ & SN = 99.60\% , TN = 99.60\% & I-FGSM, $\varepsilon$ = 0.01 & 28.62 & 5.99 & 23.25 & 1.0000 & 0.0000 \TBstrut\\ \hline
 \rowcolor{grey}
$N_{SPRITZ2}^{DGA}$ & $N_{SPRITZ1}^{N-BaIoT}$ & SN = 99.60\% , TN = 99.60\% & I-FGSM, $\varepsilon$ = 0.001 & 28.92 & 5.76 & 22.64 & 1.0000 & 0.0000 \TBstrut\\ \hline

$N_{SPRITZ2}^{DGA}$ & $N_{SPRITZ1}^{N-BaIoT}$ & SN = 99.60\% , TN = 99.60\% & JSMA, $\theta$ = 0.1 & 31.17 & 1.28 & 56.56 & 1.0000 & 0.0000  \TBstrut\\ \hline
 \rowcolor{grey}
$N_{SPRITZ2}^{DGA}$ & $N_{SPRITZ1}^{N-BaIoT}$ & SN = 99.60\% , TN = 99.60\% & JSMA, $\theta$ = 0.01 & --     & --    & --     & FAIL\textsuperscript{*}   & --      \TBstrut\\ \hline

$N_{SPRITZ2}^{DGA}$ & $N_{SPRITZ1}^{N-BaIoT}$ & SN = 99.60\% , TN = 99.60\% & LBFGS, default parameter & 34.22 & 1.40 & 33.93 & 0.9860 & 0.0040 \TBstrut\\ \hline
 \rowcolor{grey}
$N_{SPRITZ2}^{DGA}$ & $N_{SPRITZ1}^{N-BaIoT}$ & SN = 99.60\% , TN = 99.60\% & PGD, default parameter & 29.24 & 5.99 & 22.24 & 1.0000 & 0.0000 \TBstrut\\ \hline

$N_{SPRITZ2}^{DGA}$ & $N_{SPRITZ1}^{N-BaIoT}$ & SN = 99.60\% , TN = 99.60\% & DeepFool, default parameter & 31.02 & 2.59 & 53.42 & 1.0000 & 0.0104 \TBstrut\\ \hline

 \rowcolor{grey}
$N_{SPRITZ1}^{N-BaIoT}$ & $N_{SPRITZ2}^{RIPE}$ & SN = 98.80\% , TN = 99.80\% & I-FGSM, $\varepsilon$ = 0.1 & 18.65 & 22.07 & 64.78  & 1.0000 & 0.0000 \TBstrut\\ \hline

$N_{SPRITZ1}^{N-BaIoT}$ & $N_{SPRITZ2}^{RIPE}$ & SN = 98.80\% , TN = 99.80\% & JSMA, $\theta$ = 0.1 & 18.49 & 7.29  & 178.50 & 1.0000 & 0.0010 \TBstrut\\ \hline

 \rowcolor{grey}
$N_{SPRITZ1}^{N-BaIoT}$ & $N_{SPRITZ2}^{RIPE}$ & SN = 98.80\% , TN = 99.80\% & LBFGS, default parameter & 22.57 & 10.46 & 110.67 & 0.9910 & 0.0000 \TBstrut\\ \hline

$N_{SPRITZ1}^{N-BaIoT}$ & $N_{SPRITZ2}^{RIPE}$ & SN = 98.80\% , TN = 99.80\% & PGD, default parameter & 19.59 & 19.71 & 54.00  & 1.0000 & 0.0000 \TBstrut\\ \hline

\rowcolor{grey}
$N_{SPRITZ1}^{N-BaIoT}$ & $N_{SPRITZ2}^{RIPE}$ & SN = 98.80\% , TN = 99.80\% & DeepFool, default parameter & 20.87 & 12.14 & 143.19 & 0.9676 & 0.0200 \TBstrut\\ \hline

\end{tabular}}
\\FAIL\textsuperscript{*}: In this case, the attack on the source network fails and we cannot transfer it to the target network. Therefore, the numerical values of the average PSNR, the average $L_1$ dist, and the average maximum distortion are not possible. 
\end{table*}

%% file: Tables/DefenseMPA.tex
\begin{table*}[!ht]
\centering
\caption{Test Accuracy Values TN with an MPA fine-tunning. \label{Defense_MPA}}
\begin{tabular}{ | c | c | c | c | c | c | c |}
\hline
\textbf{TN} & \textbf{Fine-Tune} & \textbf{Parameter} & \textbf{I-FGSM, $\varepsilon$ = 0.01} & \textbf{IFGSM, $\varepsilon$ = 0.001} &\textbf{LBFGS, Default} &\textbf{JSMA, $\theta$ = 0.01}\TBstrut\\
 \hline
 \rowcolor{grey}
 
$N_{SPRITZ1}^{N-BaIoT}$ & JSMA & $\theta$ = 0.01 & 0.0000 & 0.0100 & 0.2010 & ---\TBstrut\\ \hline

$N_{SPRITZ1}^{DGA}$ & I-FGSM& $\varepsilon$ = 0.01 & --- & 0.0000 & 0.0010 & 0.0000\TBstrut\\ \hline
 \rowcolor{grey}

$N_{SPRITZ1}^{DGA}$ & I-FGSM& $\varepsilon$ = 0.001 & 0.0000 & --- & 0.0010 & 0.0000\TBstrut\\ \hline

$N_{SPRITZ1}^{DGA}$ & LBFGS& Default & 0.0200 & 0.0010 & --- & 0.0000\TBstrut\\ \hline

\end{tabular}
\end{table*}

%% file: Tables/executionTime.tex
\begin{table*}[!th]
\centering
\caption{Execution running attack time on SN in seconds. \label{Exe_Time}}
\resizebox{2\columnwidth}{!}{
\begin{tabular}{| c | c | c | c | c | c | c | c|}
\hline
\textbf{SN} & \textbf{Attack Type} & \textbf{Dataset1} & \textbf{Dataset2} & \textbf{Dataset3}& \{\textbf{Exe. seconds (Dataset1)} & \textbf{Exe. seconds (Dataset2)} &\textbf{Exe. seconds (Dataset3)}\TBstrut\\
 \hline
 \rowcolor{grey}
 
SPRITZ1 & I-FGSM, $\varepsilon$ = 0.1 & N-BaIoT & DGA&RIPE & 127.2 &  127.8 & 130.8 \TBstrut\\ \hline

SPRITZ1 & I-FGSM, $\varepsilon$ = 0.01 & N-BaIoT & DGA&RIPE & 1229.4 & 1294.2 & ---\TBstrut\\ \hline
 \rowcolor{grey}
SPRITZ1 & I-FGSM, $\varepsilon$ = 0.001 & N-BaIoT &DGA&RIPE &  12238.8 & 12766.2 & ---\TBstrut\\ \hline

SPRITZ1 & JSMA, $\theta$ = 0.1 & N-BaIoT &DGA&RIPE & 1824.6 & 625.2 & 1840.2 \TBstrut\\ \hline
 \rowcolor{grey}
SPRITZ1 & LBFGS, default parameter & N-BaIoT &DGA&RIPE &  2923.8 & 1555.2 & 2455.2\TBstrut\\ \hline

SPRITZ1 & PGD, default parameter & N-BaIoT &DGA&RIPE & 1015.2 & 970.2 & 976.2 \TBstrut\\ \hline
 \rowcolor{grey}
SPRITZ1 & DeepFool, default parameter & N-BaIoT &DGA&RIPE & 29.4 & 13.8 & 22.8 \TBstrut\\ \hline

SPRITZ2 & I-FGSM, $\varepsilon$ = 0.1 & N-BaIoT & DGA& RIPE & 265.2 & 269.4 & ---\TBstrut\\ \hline
\rowcolor{grey}
SPRITZ2 & I-FGSM, $\varepsilon$ = 0.01 & N-BaIoT & DGA&RIPE & 2636.4 & 2698.8 & ---\TBstrut\\ \hline
 
SPRITZ2 & I-FGSM, $\varepsilon$ = 0.001 & N-BaIoT &DGA&RIPE & 26306.16 & 26818.14 & ---\TBstrut\\ \hline
\rowcolor{grey}
SPRITZ2 & JSMA, $\theta$ = 0.1 & N-BaIoT &DGA&RIPE & 1954.8 & 1231.8 & ---\TBstrut\\ \hline
 
SPRITZ2 & LBFGS, default parameter & N-BaIoT &DGA&RIPE & 4723.8 & 3115.8 & ---\TBstrut\\ \hline
\rowcolor{grey}
SPRITZ2 & PGD, default parameter & N-BaIoT &DGA&RIPE & 2715 & 2016 & ---\TBstrut\\ \hline
 
SPRITZ2 & DeepFool, default parameter & N-BaIoT &DGA&RIPE & 18.6 & 22.2 & ---\TBstrut\\ \hline

\end{tabular}
}
\end{table*}

%% file: Conclusion.tex
\section{Conclusions and Future Work}
\label{Conclusions}
In this paper, we thoroughly dissected the transferability of adversarial examples in computer networks. We investigated three transferability scenarios: the cross training, the cross model, and the cross model and training. Considering three well-known datasets (N-BaIoT, DGA, and RIPE Atlas), we found that the transferability happens in specific cases where the adversary can perfectly transfer the examples from the SN to the TN. Although in other cases, such as the Cross Model and Training transferability, it is difficult to transfer the adversarial examples from the SN to the TN. The adversary still leverages the existing cases where the transferability happens. 
Due to the black-box nature of deep neural networks, our understanding is very narrow regarding the decisions enabling the transferability between the CNN-based models. This knowledge gap is a critical issue, especially for security-oriented applications in computer networks. Therefore, additional research is required to understand what are the main factors responsible for the transferability property. 

 According to our experiments, the JSMA and I-FGSM attacks are more transferable than other adversarial attacks, particularly in the cross model transferability. One probable rationale is that when JSMA and I-FGSM attacks alter particular features (those to which the network is more vulnerable), it is likely to overfit the attacked model. It is also worth noting that the transferability is not symmetric in terms of the training datasets for a particular task. These results indicate that the underlying dataset may influence the network's learned features in some way and to some extent in computer network applications. Thus, we need to investigate this area in future work to understand why computer networks are less vulnerable to attack transferability.
 Our future work will focus on improving the power of adversarial attacks to increase their transferability. However, this is not a straightforward process, considering the complexity of the decision boundary learned by CNNs or even other deep neural networks. 

%% file: Acknowledgment.tex
\section*{Acknowledgments}
\label{Acknowledgments}
This work is funded by the University of Padua, Italy, under the STARS Grants program (Acronym and title of the project: LIGHTHOUSE: Securing the Transition Toward the
Future Internet).